\documentclass[%
superscriptaddress,
twocolumn,
amsmath,amssymb,
aps,
pra,
]{revtex4-2}

\usepackage{graphicx}
\usepackage{dcolumn}
\usepackage{bm}
\graphicspath{{images/}{../images/}}

\begin{document}

\title{Practical continuous-variable quantum key distribution with composable security
}

\author{Nitin Jain}
\email{nitinj@iitbombay.org}
\affiliation{\mbox{Center for Macroscopic Quantum States (bigQ), Department of Physics,} Technical University of Denmark, 2800 Kongens Lyngby, Denmark}
\author{Hou-Man Chin}
\affiliation{\mbox{Department of Photonics, Technical University of Denmark}, 2800 Kongens Lyngby, Denmark}
\affiliation{\mbox{Center for Macroscopic Quantum States (bigQ), Department of Physics,} Technical University of Denmark, 2800 Kongens Lyngby, Denmark}
\author{Hossein Mani}
\affiliation{\mbox{Center for Macroscopic Quantum States (bigQ), Department of Physics,} Technical University of Denmark, 2800 Kongens Lyngby, Denmark}
\author{\mbox{Cosmo Lupo}} 
\affiliation{\mbox{Department of Physics and Astronomy,} University of Sheffield, S3 7RH Sheffield, UK}
\affiliation{\mbox{Department of Computer Science,} University of York, York YO10 5GH, UK}
\author{\mbox{Dino Solar Nikolic}}
\affiliation{\mbox{Center for Macroscopic Quantum States (bigQ), Department of Physics,} Technical University of Denmark, 2800 Kongens Lyngby, Denmark}
\author{Arne Kordts}
\affiliation{\mbox{Center for Macroscopic Quantum States (bigQ), Department of Physics,} Technical University of Denmark, 2800 Kongens Lyngby, Denmark}
\author{Stefano Pirandola} 
\affiliation{\mbox{Department of Computer Science,} University of York, York YO10 5GH, UK}
\author{Thomas Brochmann Pedersen}
\affiliation{Cryptomathic A/S, Aaboulevarden 22, 8000 Aarhus, Denmark}
\author{\mbox{Matthias Kolb}} 
\author{\mbox{Bernhard {\"O}mer}} 
\author{Christoph Pacher} 
\affiliation{Center for Digital Safety \& Security, AIT Austrian Institute of Technology GmbH, 1210 Vienna, Austria.}
\author{Tobias Gehring}
\email{tobias.gehring@fysik.dtu.dk}
\author{Ulrik L. Andersen}
\email{ulrik.andersen@fysik.dtu.dk}
\affiliation{\mbox{Center for Macroscopic Quantum States (bigQ), Department of Physics,} Technical University of Denmark, 2800 Kongens Lyngby, Denmark}

\date{\today}

\begin{abstract}
A quantum key distribution (QKD) system must fulfill the requirement of universal composability to ensure that any cryptographic application (using the QKD system) is also secure. Furthermore, the theoretical proof responsible for security analysis and key generation should cater to the number $N$ of the distributed quantum states being finite in practice. Continuous-variable (CV) QKD based on coherent states, despite being a suitable candidate for integration in the telecom infrastructure, has so far been unable to demonstrate composability as existing proofs require a rather large $N$ for successful key generation. Here we report the first Gaussian-modulated coherent state CVQKD system that is able to overcome these challenges and can generate composable keys secure against collective attacks with $N \lesssim 3.5\times10^8$ coherent states. With this advance, possible due to novel improvements to the security proof and a fast, yet low-noise and highly stable system operation, CVQKD implementations take a significant step towards their discrete-variable counterparts in practicality, performance, and security.

\begin{description}
\item[PACS numbers]
May be entered using the \verb+\pacs{#1}+ command.
\end{description}
\end{abstract}

\pacs{Valid PACS appear here}
\maketitle

\section{Introduction}
Quantum key distribution (QKD) is the only known cryptographic solution for distributing secret keys to users across a public communication channel while being able to detect the presence of an eavesdropper~\cite{Scarani2009, Pirandola2019}. Legitimate QKD users (Alice and Bob) encrypt their messages with the secret keys and exchange them with the assurance that the eavesdropper (Eve) cannot break the confidentiality of the encrypted messages. In particular, if the obtained secret key is (at least) as long as the length of the message, information theoretic security guarantees that Eve cannot break the security even if equipped with unlimited computing resources. 

\begin{figure*}
    \centering
    \includegraphics[width=0.98\linewidth]{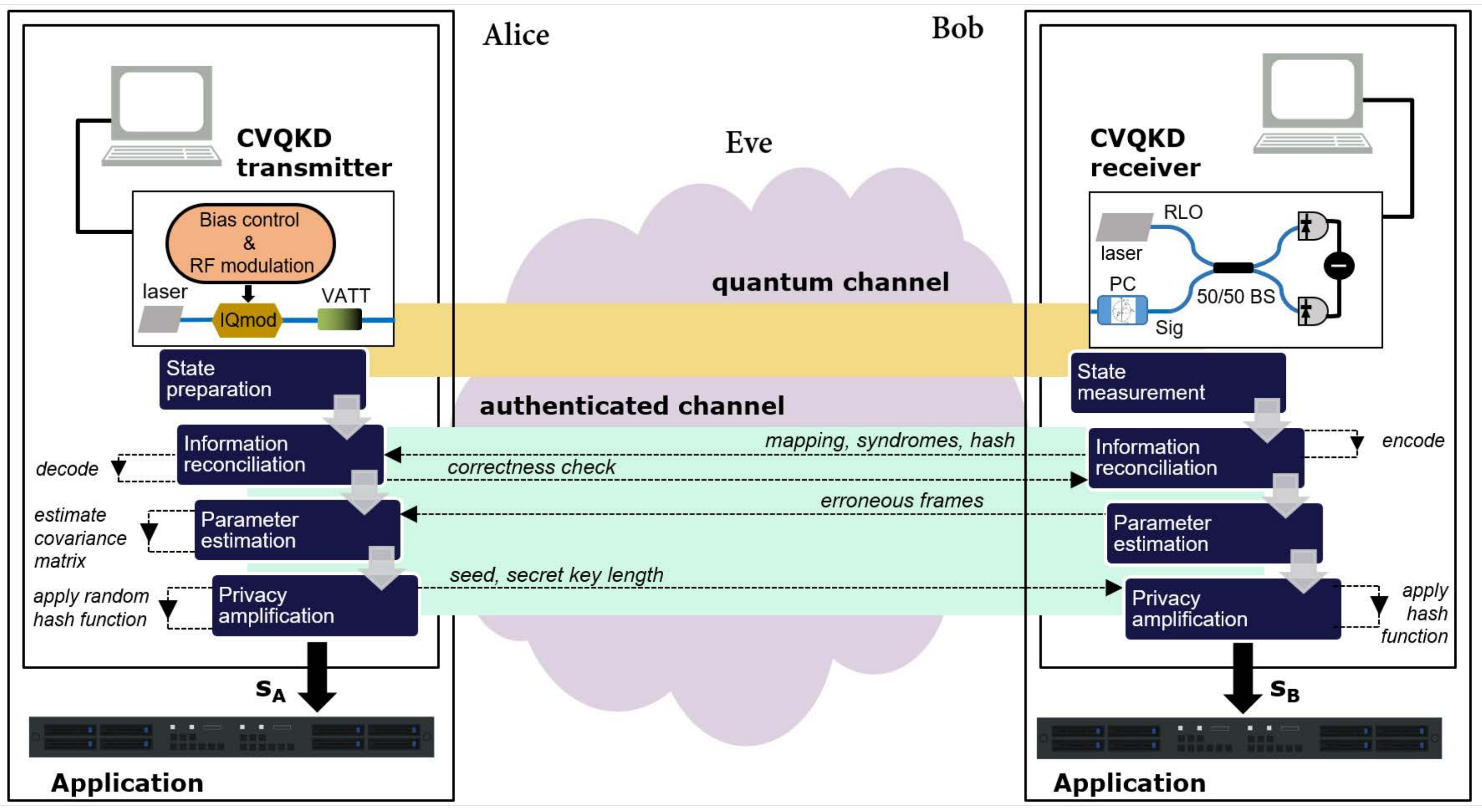}
    \caption{Composability in continuous-variable quantum key distribution (CVQKD). Alice and Bob obtain bitstreams $s_A$ and $s_B$, respectively, after going through the different steps of the QKD protocol that involve both the quantum and authenticated channels, assumed to be in full control of Eve. Various dashed lines (with arrows) indicate local operations and classical communication. An application using a CVQKD system (transmitter and receiver) to provide composable security must satisfy certain criteria associated with correctness, robustness, and secrecy of the protocol~\cite{Muller-Quade2009, Leverrier2015}. For instance, the condition Pr$[s_A \neq s_B] \leq \epsilon_{\rm cor}$ denotes $\epsilon$-correctness, and describes the failure probability of Alice and Bob having non-identical keys at the conclusion of the protocol. Details of our CVQKD implementation (including full forms of the acronyms used in this  graphic) are presented in \ref{Exp:Tx} and \ref{Exp:Rx}, and the details of the protocol operation can be found in \ref{Exp:Protocol}.} 
    \label{fig:scheme}
\end{figure*}

Alice and Bob perform a sequence of steps, shown in Fig.~\ref{fig:scheme}, to obtain a key of a certain length. Such a `QKD protocol' begins with preparation, transmission (on a quantum channel), measurement of quantum states, and concludes with classical data processing and security analysis, performed in accordance with a mathematical proof. 
Amongst the many physical considerations included in the security proof, one is that the number of quantum states available to Alice and Bob are not infinite. Such \textit{finite-size corrections} adversely affect the key length but are essential for the security assurance. 
 
Another related property of a cryptographic key is \textit{composability}~\cite{Canetti2001}, which allows specifying the security requirements for combining different cryptographic applications in a unified and systematic way. In the context of practical QKD, composability is of utmost importance because the secret keys obtained from a QKD protocol are almost always used in other applications, e.g.\ data encryption~\cite{Muller-Quade2009}. A QKD implementation that outputs a key not proven to be composable is thus practically useless. 

In one of the most well-known flavours of QKD, the quantum information is coded in continuous variables, such as the amplitude and phase quadratures, of the optical field~\cite{Ralph99, Diamanti2015, Laudenbach2018, Pirandola2019}. Typical continuous-variable (CV)QKD protocols have been Gaussian-modulated coherent state (GMCS) implementations~\cite{Jouguet2013, Huang2016, Wang2018, Wang2020}, and finite-size effects were also considered, though the proof~\cite{Leverrier2010} was non-composable. Composable security in CVQKD was first proven and experimentally demonstrated using two-mode squeezed states, however, since the employed entropic uncertainty relation is not tight, the achievable communication distance was rather limited~\cite{Furrer2012, Gehring2015}. 

Composable security proofs for CVQKD systems using coherent states and dual quadrature detection, first proposed in 2015~\cite{Leverrier2015}, have been progressively improved~\cite{Lupo2018, Papanastasiou2021, Pirandola2021}. These proofs promise keys at distances much longer than in Ref.~\cite{Furrer2012} apart from the advantage of dealing with coherent states, which are much easier to generate than squeezed states. Nonetheless, an experimental demonstration of composability has remained elusive, due to a combination of the strict security bounds (because of a complex parameter estimation routine), the large number of required quantum state transmissions (to keep the finite-size terms sufficiently low), and the stringent requirements on the tolerable excess noise. 

In this article, we demonstrate a practical GMCS-CVQKD system that is capable of generating composable keys secure against collective attacks. We achieve this by deriving a new method for establishing confidence intervals that is compatible with collective attacks, which allows us to work on smaller (and thus more practical) block sizes than originally required~\cite{Leverrier2015}. On the experimental front, we are able to keep the excess noise below the null key length threshold by performing a careful analysis (followed by eradication or avoidance) of the various spurious noise components, and by implementing a machine learning framework for phase compensation~\cite{Chin2020}. 

After taking finite-size effects as well as confidence intervals from various system calibrations into account, we achieve a positive composable key length with merely $N \lesssim 3.5\times10^8$ coherent states (also referred to as `quantum symbols'). With $N = 10^9$, we obtain $>53.4$ Mbits worth of composably secure key material in the worst case. 

\section{Composably secure key}\label{sec:Theory}
In the security analysis, we assume collective attacks and take into account the finite number of coherent states transmitted by Alice and measured by Bob. A digital signal processing (DSP) routine yields the digital quantum symbols $\bar{Y}$ discretized with $d$ bits per quadrature and this stream is divided into $M$ frames for information reconciliation (IR), after which we perform parameter estimation (PE) and privacy amplification (PA); as visualized in Fig.~\ref{fig:scheme}. We derive the secret key bound for reverse reconciliation, i.e., Alice correcting her data according to Bob's quantum symbols. 

The (composable) secret key length $s_n$ for $n$ coherent state transmissions is calculated using tools from Refs.~\cite{Leverrier2015, Pirandola2021} as well as new results presented in the following. The key length is bounded per the leftover hash lemma in terms of the smooth min-entropy $H_{\min}$ of the alphabet of $\bar{Y}$, conditioned on the quantum state of the eavesdropper $E$~\cite{Tomamichel2012}. From this we subtract the information reconciliation leakage $\mathrm{leak_{IR}}(n,\epsilon_{\mathrm{IR}})$ and obtain, 
\begin{equation}
s_{n}^{\epsilon_h+\epsilon_s+\epsilon_{\mathrm{IR}}} 
\geq 
H_{\min}^{\epsilon_s}(\bar Y | E )_{ \rho^{n} } -\mathrm{leak_{IR}}(n,\epsilon_{\mathrm{IR}}) 
+ 2 \log_2{(\sqrt{2}\epsilon_h)} \, .
\label{eq:skfSimple}
\end{equation}
The security parameter $\epsilon_h$ characterizes the hashing function, $\epsilon_s$ is the smoothing parameter entering the smooth conditional min-entropy, and $\epsilon_{\mathrm{IR}}$ describes the failure probability of the correctness test after IR. 

The probability $p^\prime$ that IR succeeds in a frame is related to the frame error rate (FER) by $p^\prime = 1 - $FER. All frames in which IR failed are discarded from the raw key stream, and this step thereby projects the original tensor product state $\rho^{n} \equiv \rho^{\otimes n}$ into a non i.i.d. state $\tau^{n}$. To take this into account, one replaces the smooth min-entropy term in Eq.~\eqref{eq:skfSimple} with the expression~\cite{ Pirandola2021}: 
\begin{align}\label{new-min}
H_{\min}^{\epsilon_s}(\bar Y | E)_{\tau^{n'}} 
\geq H_{\min}^{\frac{p^\prime}{3} \epsilon_s^2}(\bar Y |E)_{ \rho^{\otimes n'} } 
+ \log_2{\left(  p^\prime - \frac{p^\prime}{3}\epsilon_s^2\right)  } \, ,
\end{align}
where $n' = n p^\prime$ is the number of quantum symbols remaining after error correction.

The asymptotic equipartition property (AEP) bounds the conditional min-entropy by the von-Neumann conditional entropy,
\begin{equation*}
H_{\min}^{\delta}(\bar Y | E)_{\rho^{\otimes n'}}
\geq n' H(\bar Y |E)_{\rho}-\sqrt{n'}\,\Delta_{\mathrm{AEP}}(\delta,d)\ ,
\end{equation*}
where 
\begin{equation}
\Delta_{\mathrm{AEP}}(\delta,d)\leq 4(d+1)\sqrt{\log_2{(2/\delta^{2})}}\ ,
\end{equation}
is an improved penalty in comparison to Ref.~\cite{Leverrier2015,Pirandola2021}, and is proven in the Supplement. 

The conditional von-Neumann entropy is given by
\begin{equation}
H(\bar Y | E )_\rho = H(\bar Y)_\rho - I(\bar Y ; E)_\rho \, .
\end{equation} 

We estimate the first term directly from the data (up to a probability not larger than $\epsilon_\mathrm{ent}$; further details regarding the confidence intervals are in the Supplement). The second term is bound by the Holevo information,
\begin{equation*}
I(\bar Y ; E)_\rho \leq I( Y ; E )_\rho \leq I( Y ; E )_{\rho_G} \ ,
\end{equation*}
where $Y$ is the continuous version of $\bar Y$ and $I( Y ; E )_{\rho_G}$ is the Holevo information obtained after using the extremality property of Gaussian attacks. 

The Holevo information is estimated by evaluating the covariance matrix using worst-case estimates for its entries based on confidence intervals. We improved the confidence intervals of Ref.~\cite{Leverrier2015} by exploiting the properties of the beta distribution. Let $\hat{x}$, $\hat{y}$, $\hat{z}$ be the estimators for the variance of the transmitted ensemble of coherent states, the received variance and the co-variance, respectively. The true values $y$ and $z$ are bound by
\begin{align}
    y &\le \left(1 + \delta_\text{Var}(n, \epsilon_\text{PE}/2)\right)\hat{y}\ , \label{eq:dvar} \\
    z &\ge \left(1 - 2\delta_\text{Cov}(n, \epsilon_\text{PE}/2)\frac{\sqrt{\hat x \hat y}}{\hat z}\right)\hat{z}\ , \label{eq:dcov}
\end{align}
with $\epsilon_\text{PE}$ denoting the failure probability of parameter estimation, and
\begin{align*}
\delta_\mathrm{Var} (n,\epsilon) & = a'\left(\epsilon/6\right) \left( 1 + \frac{120}{\epsilon} e^{-\frac{n}{16}} \right) - 1 \ ,  \\
\delta_\mathrm{Cov}(n,\epsilon) & =  \frac{1}{2}\left[\frac{a'\left(\epsilon/6\right)-b'\left(\epsilon/6\right)}{2}
+ a'\left(\frac{\epsilon^2}{324}\right)
- b'\left(\frac{\epsilon^2}{324} \right)
\right]
\end{align*}
being the new confidence intervals (derived in the Supplement). In the above equations, 
\begin{align*}
a'\left(\epsilon\right) & = 2 \left[ 1 - \mathrm{invcdf}_{\mathrm{Beta}(n/2,n/2)}\left(\epsilon\right)  \right] \, , \\
b'\left(\epsilon\right) & = 2 \, \mathrm{invcdf}_{\mathrm{Beta}(n/2,n/2)}\left(\epsilon\right)  \, .
\end{align*}
As detailed in section~\ref{sec:resNdis}, the (length of the) secret key we eventually obtain in our experiment requires an order of magnitude lower $N$ due to these confidence intervals. 

Finally, we remark here on a technical limitation arising due to the digitization of Alice's and Bob's data. In practice, it is impossible to implement a \emph{true} GMCS protocol because the Gaussian distribution is both unbounded and continuous, while the devices used in typical CVQKD systems have a finite range and bit resolution~\cite{Jouguet2012}. In our work, we consider a range of 7 standard deviations and use $d=6$ bits (leading to a constellation with $2^{2d} = 4096$ coherent states), which per recent results~\cite{Lupo2020, Denys2021}, should suffice to minimise the impact of digitization on the security of the protocol.
\section{Experiment}
\begin{figure*}
\centering
\includegraphics[width=0.98\linewidth]{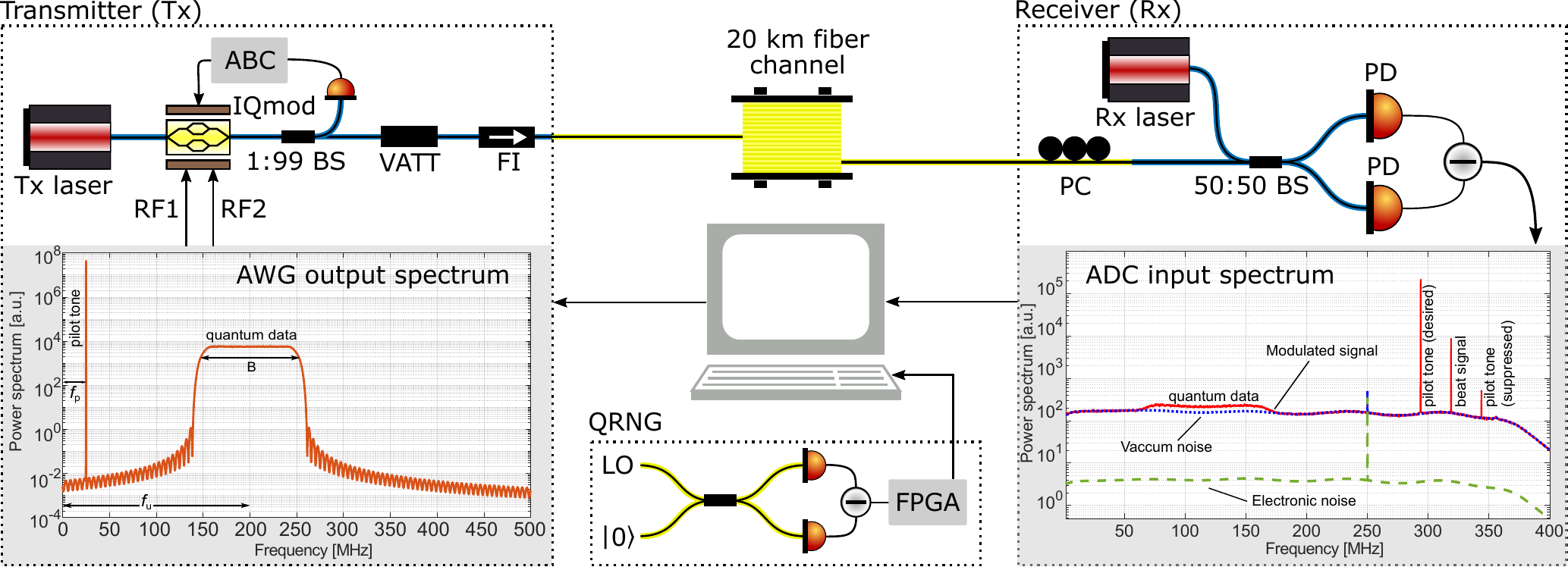}
\caption{\textbf{Schematic of the experiment.} The transmitter (Tx) and receiver (Rx) were built from polarization maintaining fiber components. The transmitter comprised a 1550\,nm continuous-wave laser (Tx laser), an in-phase and quadrature electro-optic modulator (IQmod) with automatic bias controller (ABC) for carrier suppression and single sideband modulation, and a variable attenuator (VATT) and faraday isolator (FI). An arbitrary waveform generator (AWG) with 16 bit resolution and sampling rate of 1 GSps supplied waveforms RF1 and RF2 for driving IQmod. A quantum random number generator (QRNG) delivered Gaussian-distributed symbols for discrete Gaussian modulation of coherent states. The receiver comprised a laser (Rx laser; same type as Tx laser), a polarization controller (PC) to tune the incoming signal field's polarization, a symmetric beam splitter followed by a balanced detector for radio-frequency heterodyne detection. The detector's output was sampled by a 16 bit analog-to-digital converter (ADC) at 1 GSps. BS: beam splitter, PD: photo detector. Left inset: Power spectrum of the complex waveform RF1 + $\iota$ RF2 driving the IQmod. Right inset: Power spectra of the receiver from 3 different measurements described in section~\ref{Exp:Rx}. The noise peak at 250 MHz is an interleaving spur of the ADC. 
\label{fig:setup}}
\end{figure*}
Figure~\ref{fig:setup} shows the schematic of our setup, with the caption detailing the components and their role briefly. Below we summarize the setup's operation, calibration measurements, and our protocol implementation. In the Supplement, we describe the different functional blocks of Fig.~\ref{fig:setup} in further detail. 
\subsection{Transmitter (Tx)}\label{Exp:Tx}
We performed optical single sideband modulation with carrier suppression (OSSB-CS) using an off-the-shelf IQ modulator and automatic bias controller (ABC). An arbitrary waveform generator (AWG) was connected to the RF ports to modulate the sidebands. The coherent states were produced in a $B=100\,$MHz wide frequency sideband, shifted away from the optical carrier~\cite{Lance2005, Jain2021}. The random numbers that formed the complex amplitudes of these coherent states were drawn from a Gaussian distribution, obtained by transforming the uniform distribution of a vacuum-fluctuation based quantum random number generator (QRNG), with a security parameter $\epsilon_\text{qrng} = 2 \times 10^{-6}$~\cite{Gehring2021}.

To this wideband `quantum data' signal, centered at $f_u=200\,$MHz, we multiplexed in frequency a `pilot tone' at $f_p=25\,$MHz for sharing a phase reference with the receiver~\cite{Qi2015, Soh2015, Huang2015, Kleis2017}. The left inset of Fig.~\ref{fig:setup} shows the complex spectra of the RF modulation signal.
\subsection{Receiver (Rx)} \label{Exp:Rx}
After propagating through the quantum channel---a 20 km long standard single mode fiber spool---the signal field's polarization was manually tuned to match the polarization of the real local oscillator (RLO) for heterodyning~\cite{Qi2015,Soh2015,Huang2015}. The Rx laser that supplied the RLO was free-running with respect to the Tx laser and detuned in frequency by $\sim320\,$ MHz, giving rise to a beat signal, as labelled in the solid-red spectral trace in the right inset of Fig.~\ref{fig:setup}. The quantum data band and pilot tone generated by the AWG are also labelled. Due to finite OSSB~\cite{Jain2021}, a suppressed pilot tone is also visible; the corresponding suppressed quantum band was however outside the receiver bandwidth (we used a low pass filter with a cutoff frequency around 360 MHz at the output of the heterodyne detector). 

In separate measurements, we also measured the vacuum noise (Tx laser off, Rx laser on) and the electronic noise of the detector (both Tx and Rx lasers off), depicted by the dotted-blue and dashed-green traces, respectively, in the right inset of Fig.~\ref{fig:setup}. The clearance of the vacuum noise over the electronic noise is $>15\,$dB over the entire quantum data band. 
\subsection{Noise analysis \& Calibration}\label{Exp:NoiseCal}
\begin{table*}[!t]
\centering
\caption{Experimental parameters. PNU: photon number units. The security parameter $\epsilon_\text{qrng}$ is limited by the digitization error of the ADC used in the QRNG, but could be improved using longer measurement periods~\cite{Gehring2021}.}
\begin{tabular}{l|l}
    \textbf{Transmitter} & \\
    \hline
    Rate of coherent states, $B$ & 100 MSymbols/s \\
    Modulation strength (channel input), $\mu$ & 1.45 PNU \\
    \hline
    \textbf{Receiver calibration} & \\
    \hline
    Trusted efficiency (incl. optical loss), $\tau$ & 0.69 \\
    Trusted electronic noise, $t$ & 25.71$\,\times 10^{-3}$ PNU \\
    \hline
    \textbf{Channel parameter estimation} & \\
    \hline
    Untrusted efficiency, $\eta$ & 0.35 \\
    Untrusted excess noise, $u$ & 6.30$\,\times 10^{-3}$ PNU \\
    \hline
    \textbf{Information reconciliation} & \\
    \hline
    Signal-to-noise ratio & 0.32 \\
    Frame error rate, FER & 0.36\% \\
    Reconciliation efficiency, $\beta$ & 91.6\% \\    
    Leaked bits & $1.60 \times 10^9$ \\
    \hline
    \textbf{Secret key calculation} & \\
    \hline
    Raw key length (symbols), $N_\text{PA}$ & $9.84 \times 10^8$ \\
    Security parameters & $\epsilon_h = \epsilon_\text{cal} = \epsilon_s = \epsilon_\text{PE} = 10^{-10}$, $ \epsilon_\text{qrng} = 2 \times 10^{-6}$,  $\epsilon_\text{IR} = 10^{-12}$ \\
    Final secret key length (bits) & 53452436 \\  
\end{tabular}
\label{tab:exp_params}
\end{table*}
A careful choice of the parameters defining the pilot tone and the quantum data band, and their locations with respect to the beat signal is crucial in minimizing the excess noise. A strong pilot tone enables more accurate phase reference but at the expense of higher leakage in the quantum band and an increased number of spurious tones. The latter may arise as a result of frequency mixing of the (desired) pilot tone with e.g., the beat signal or the suppressed pilot tone. As can be observed in the right inset of Fig.~\ref{fig:setup}, we avoided spurious noise peaks  resulting from sum- or difference-frequency generation of the various discrete components (in the solid-red trace) from landing inside the wide quantum data band. 

As is well known in CVQKD implementations, Alice needs to optimize the modulation strength of the coherent state alphabet at the input of the quantum channel to maximize the secret key length. For this, we connected the transmitter and receiver directly, i.e., without the quantum channel, and performed heterodyne measurements to calibrate the \emph{mean photon number} $\mu$ of the resulting thermal state from the ensemble of generated coherent states, as explained in section~\ref{Exp:Tx}. The modulation strength can be controlled in a fine-grained manner using the electronic gain of the AWG and the optical attenuation from the VATT. The DSP that aided in this calibration is explained in detail in the supplement. 

Since we conducted our experiment in the non-paranoid scenario~\cite{Scarani2009, Jouguet2012}, i.e., we trusted some parts of the overall loss and excess noise by assuming them to be beyond Eve's control, some extra measurements and calibrations for the estimation of trusted parameters become necessary. More specifically, we decomposed the total transmittance and excess noise into respective trusted and untrusted components. In the Supplement, we present the details of how we evaluated the trusted transmittance $\tau$ and trusted noise $t$ for our setup. 

Table~\ref{tab:exp_params} presents the values of $\mu$, $\tau$ and $t$ pertinent to the experimental measurement described in section~\ref{Exp:Protocol}. Let us remark here that in our work, we express the noise and other variance-like quantities, e.g., the modulation strength, in photon number units (PNU) as opposed to the traditional shot noise units (SNU) because the former is independent of quadratures, and in case of $\mu$, facilitates a comparison with discrete-variable (DV) QKD systems\footnote{Assuming symmetry between the quadratures, the modulation variance $V_{\text{mod}} = 2 \mu$ in SNU.}. Finally, note that we recorded a total of $10^{10}$ ADC samples for each of the calibration measurements, and all the acquired data was stored on a hard drive for offline processing. 
\subsection{Protocol operation}\label{Exp:Protocol}
We connected the transmitter and receiver using the 20 km channel, optimized the signal polarization, and then collected heterodyne data using the same Gaussian distributed random numbers as mentioned in section~\ref{Exp:NoiseCal}. Offline DSP~\cite{Chin2020} was performed at the receiver workstation to obtain the symbols that formed the raw key. The preparation and measurement was performed with a total of $10^9$ complex symbols, modulated and acquired in 25 blocks, each block containing $4 \times 10^7$ symbols. After discarding some symbols due to a synchronization delay, Alice and Bob had a total of $N_\text{IR} = 9.88 \times 10^8$ correlated symbols at the beginning of the classical phase of the protocol; see Fig.~\ref{fig:scheme}. 

Below we provide details of the actual protocol we implemented, where we assumed that the classical channel connecting Alice and Bob was already authenticated. 
\begin{enumerate}
    \item IR was based on a multi-dimensional scheme~\cite{MD-Recon-PRA.77.042325} using multi-edge-type low-density-parity-check error correcting codes~\cite{Mani2018}. Table~\ref{tab:exp_params} lists some parameters related to the operating regime and the performance of these codes; more information is available in the Supplement. As shown in Fig.~\ref{fig:scheme}, Bob sent the mapping and the syndromes, together with the hashes computed using a randomly chosen Toeplitz function, to Alice, who performed  correctness confirmation and communicated it to Bob. 
    \item During PE, Alice estimated the entropy of the corrected symbols, and together with the symbols from the erroneous frames, i.e., frames that could not be reconciled successfully (and were publicly announced by Bob), Alice evaluated the covariance matrix. This was followed by evaluating the channel parameters as well as performing the `parameter estimation test' (refer Theorem 2 in Ref.~\cite{Leverrier2015}) and getting a bound on Eve's Holevo information. Using the expression for the secret key length with the security parameters from Table~\ref{tab:exp_params}, Alice then calculated the number of bits expected in the output secret key in the worst-case scenario. This length was communicated together with a seed to Bob. 
    \item For PA, the shared seed from the previous step was used to select a random Toeplitz hash function by Alice and Bob, who then employed the high-speed and large-scale PA scheme~\cite{Tang2019} to generate the final secret key. 
\end{enumerate}

\section{Results \& Discussion}\label{sec:resNdis} 
\begin{figure*}[!t]
\centering
\includegraphics[width=0.98\linewidth]{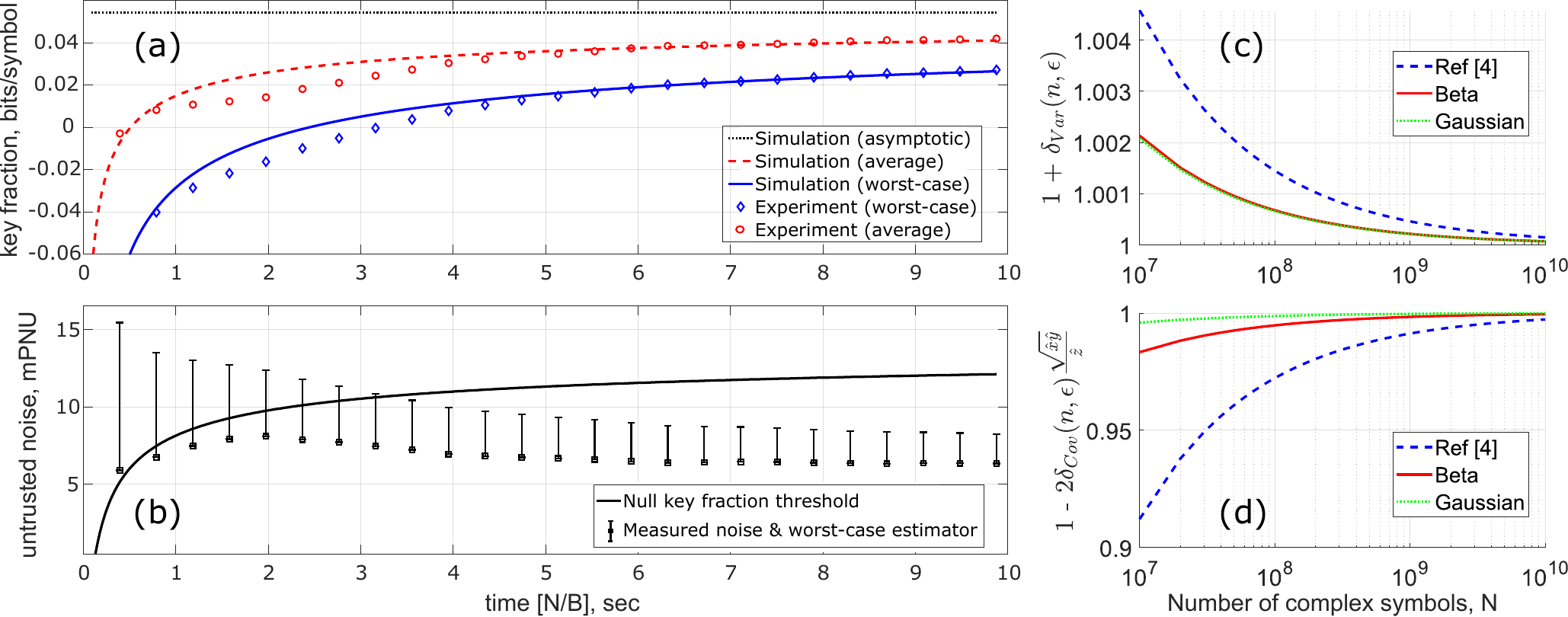}
    \caption{Composable SKF results. (a) Pseudo-temporal evolution of the composable SKF with the time parameter calculated as the ratio of the cumulative number $N$ of complex symbols available for the classical steps of the protocol and the rate $B$ at which these symbols are modulated. (b) Variation of untrusted noise $u$ measured in the experiment (lower point) and its worst-case estimator (upper point), and the noise threshold to beat in order to get a positive composable SKF. The reason for the deviation of the traces in (a) from the experimental data between 1 and 4 seconds is due to the slight increase in $u$. (c) and (d) Comparison of confidence intervals derived in this manuscript (Beta; solid-red trace and Gaussian; dashed-green trace) with those derived in the original composable security proof (Ref.~\cite{Leverrier2015}; dashed-blue trace) as a function of $N$. 
    \label{fig:results}}
\end{figure*}
Table~\ref{tab:exp_params} summarizes the relevant parameters in our experiment. Alice prepared an ensemble of $10^9$ coherent states, characterized by a modulation strength of 1.45 PNU, transmitted them over a 20 km channel to Bob, who measured them with a total excess noise $u+t = 6.3 + 25.7 = 32.0\,$mPNU and a total transmittance $\eta \cdot \tau = 0.35 \cdot 0.69 = 0.24$ averaged over the amplitude and phase (I and Q) quadratures. With a total of $N_\text{IR} = 9.88 \times 10^8$ correlated  symbols, Bob and Alice performed reverse reconciliation with an efficiency $\beta = $ 91.6\% as explained in section~\ref{Exp:Protocol}. Notably, due to the low frame error rate (FER = 0.0036) during IR, Alice and Bob were left with $N_\text{PA} = 9.84 \times 10^8$ symbols for performing the last classical step of the protocol. 

Using the equations presented in section~\ref{sec:Theory}, we can calculate the composably secure key length (in bits) for a certain number $N$ of the quantum symbols. We partitioned $N_\text{IR}$ in 25 blocks, estimated the key length considering the total number $N_k$ of symbols accumulated from the first $k$ blocks, for $k \in \{1, 2, \ldots, 25\}$. Dividing this length by $N_k$ yields the composable secret key fraction (SKF) in bits/symbols. If we neglect the time taken by data acquisition, DSP, and the classical steps of the protocol, i.e., only consider the time taken to modulate $N=N_k$ coherent states at the transmitter (at a rate $B = 100\,$MSymbols/s), we can construct a hypothetical time axis to show the evolution of the CVQKD system.

Figure~\ref{fig:results}(a) depicts such a time evolution of the SKF after proper consideration to the finite-size corrections due to the average and worst-case (red and blue data points, respectively) values of the underlying parameters. Similarly, Fig.~\ref{fig:results}(b) shows the experimentally measured untrusted noise $u$ (lower squares) together with the worst-case estimator (upper dashes) calculated using $N_k$ in the security analysis. To obtain a positive key length, the worst-case estimator must be below the maximum tolerable noise---null key fraction threshold---shown by the solid line, and this occurs at $N/B \lesssim 3.5$ seconds. 

Note that in reality, the DSP and classical data processing consume a significantly long time: In fact, we store the data from the state preparation and measurement stages on disks and perform these steps offline. The plots in Fig.~\ref{fig:results} therefore may be understood to be depicting the time evolution of the SKF and the untrusted noise \emph{if} the entire protocol operation was in real time.

Joining data from both I and Q quadratures bestowed $2 N_\text{PA} = 2 \times 9.84 \times 10^8$ \emph{real} symbols, from which we then obtain a secret key with length $l = 53452436$ bits, implying a worst-case SKF = $0.027\,$bits/symbol. Referring to Fig.~\ref{fig:results}(a), the solid-blue and dashed-red traces simulate the SKF in the worst-case and average scenarios, respectively, while the dotted-black trace shows the asymptotic SKF value obtainable with the given channel parameters; refer Table~\ref{tab:exp_params}. Per projections based on the simulation, the worst-case composable SKF should be within 1\% of the asymptotic value for $N \approx 10^{12}$ complex symbols. 

From a theoretical perspective, the reason for being able to generate a positive composable key length with a relatively small number of coherent states ($N \lesssim 3.5\times10^8$) can mainly be attributed to the improvement in confidence intervals during PE; refer equations~\ref{eq:dvar} and \ref{eq:dcov}. Figures~\ref{fig:results}(c) and (d) quantitatively compare the scaling factor in the RHS of these equations, respectively, as a function of $N$ for three different distributions. The estimators $\hat{x}$, $\hat{y}$, $\hat{z}$ for this purpose are the actual values obtained in our experiment and we used an $\epsilon_\text{PE} = 10^{-10}$. The difference between the confidence intervals used in Ref.~\cite{Leverrier2015} (suitably modified here for a fair comparison) with those derived here, based on the Beta distribution, is quite evident at lower values of $N$, as visualized by comparing the dashed-blue trace with the solid-red one. 

Since the untrusted noise has a quadratic dependence on the covariance in contrast to variance where the dependence is linear, a method that tightens the confidence intervals for the covariance can be expected to have a large impact on the final composable SKF. In fact, according to simulation, our implementation would have required almost an order of magnitude higher $N_\text{PA} $ ($\gtrsim 7.5 \times 10^9$) using the confidence intervals of Ref.~\cite{Leverrier2015} to achieve the peak SKF depicted by the rightmost blue data point in Fig.~\ref{fig:results}(a).

The dashed-green trace shows the confidence intervals also based on the Beta distribution, and a further assumption of the underlying data, i.e., the I and Q quadrature symbols, following a Gaussian distribution (more details provided in the Supplement). This however may restrict the security analysis to Gaussian collective attacks, therefore, we do not make this assumption in our calculations. The advantage of this method would however be even tighter confidence intervals, and thus, even lower requirements on $N$ for obtaining a composable key with positive length. 

On the practical front, a reasonably large transmission rate $B = 100\,$MSymbols/s of the coherent states together with the careful analysis and removal of excess noise (refer section~\ref{Exp:NoiseCal} for more details) enables an overall fast, yet low-noise and highly stable system operation, critical in quickly distributing raw correlations of high quality and keeping the finite-size corrections minimal. 

\section{Conclusion \& Outlook}
Due to its similarity to coherent telecommunication systems, continuous-variable quantum key distribution (CVQKD) based on coherent states is perhaps the most cost-effective solution for widespread deployment of quantum cryptography at access network scales ($10-50\,$km long quantum channels). However, CVQKD protocols have lagged behind their discrete-variable counterparts in terms of security, particularly, in demonstrating composability and robustness against finite-size effects. In this work, we have implemented a prepare-and-measure Gaussian-modulated coherent state CVQKD protocol that operates over a 20 km long quantum channel connecting Alice and Bob, who, at the end of the protocol obtain a composable secret key that takes finite-size effects into account and is protected against collective attacks. Our achievement was enabled by means of several novel advances in the theoretical security analysis and technical improvements on the experimental front. Furthermore, by using a real local oscillator at the receiver, we enhance the practicality as well as the security of the QKD system against hacking. 

In conclusion, we believe this is a significant advance that demonstrates practicality, performance, and security of CVQKD implementations operating in the low-to-moderate channel loss regime. With an order of magnitude larger $N$ and half the current value of $u$, we expect to obtain a non-zero length of the composable key while tolerating channel losses around 8 dB, i.e.,\ distances up to $\sim 40\,$km (assuming an attenuation factor of 0.2 dB/km). This should be easily achievable with some improvements in the hardware as well as the digital signal processing. We therefore expect that in the future, users across a point-to-point link could use the composable keys from our CVQKD implementation to enable real applications such as secure data encryption, thus ushering in a new era for CVQKD. \\

\section*{Acknowledgements}
We thank Marco Tomamichel for discussions regarding the security analysis. The work presented in this paper has been supported by the European Union's Horizon 2020 research and innovation programmes CiViQ (grant agreement no.\ 820466), OPENQKD (grant agreement no.\ 857156), and CSA Twinning NONGAUSS (grant agreement no.\ 951737). NJ, HMC, HM, ULA, and TG acknowledge support from Innovation Fund Denmark (CryptQ project, grant agreement no.\ 0175-00018A) and the Danish National Research Foundation, Center for Macroscopic Quantum States (bigQ, DNRF142). CL acknowledges funding from the EPSRC Quantum Communications Hub, Grant No. P/M013472/1 and EP/T001011/1. 

\bibliographystyle{unsrtAuthabbv} 
\bibliography{lib}

\end{document}


\maketitle

\section{Improved confidence intervals}
The goal of the parameter estimation routine is to obtain, from empirical data, an estimate, with confidence intervals, of the relevant parameters that characterize the CVQKD protocol. For our protocol it is essential to estimate the variance and covariance of the outcome of unbounded variables, e.g., during the calibration process and for the estimation of the secret key fraction. These unbounded variables are either the outputs of heterodyne detection on the receiver's side, or the preparation variables on the transmitter's side. 
In all we assume $2n$ input (amplitude and phase quadrature) values that yield us $n$ complex symbols. At the transmitter, we denote the $j$-th symbol by $(X_{2j-1},X_{2j}) := (q^j_\text{tx},p^j_\text{tx})$, and correspondingly, the receiver obtains the symbol $(Y_{2j-1},Y_{2j}) := (q^j_\text{rx},p^j_\text{rx})$ through heterodyne measurements. Under the assumption of symmetry of the quadrature variables, we consider the empirical variance and covariance: 
\begin{align}
\hat{x} &:= \mathrm{Var}(X) := \frac{1}{2n}\| X \|^2
 := \frac{1}{2n} \sum_{j=1}^{2n} X_j^2 = \frac{1}{n} \sum_{j=1}^{n} \frac{\left(q^j_\text{tx}\right)^2 + \left(p^j_\text{tx}\right)^2}{2} \ , \label{eqs:cvmX} \\
\hat{y} &:= \mathrm{Var}(Y) := \frac{1}{2n}\| Y \|^2
 := \frac{1}{2n} \sum_{j=1}^{2n} Y_j^2 = \frac{1}{n} \sum_{j=1}^{n} \frac{ \left(q^j_\text{rx}\right)^2 + \left(p^j_\text{rx}\right)^2 }{2} \ , \label{eqs:cvmY} \\
\hat{z} &:= \mathrm{Cov}(X,Y) := \frac{1}{2n}\langle X, Y \rangle 
 := \frac{1}{2n} \sum_{j=1}^{2n} X_j Y_j = \frac{1}{n} \sum_{j=1}^{n} \left( \frac{ q^j_\text{tx} q^j_\text{rx} + p^j_\text{tx} p^j_\text{rx} }{2} \right) \ .
 \label{eqs:cvmZ}
\end{align}

A rigorous calculation of confidence intervals for these variables was performed in Ref.~\cite{Leverrier2015}. Here we follow this approach and derive improved bounds. In particular, we modify Lemma~7 to deal with numerical values instead of the exponential bounds of the $\chi^2$-distribution.

\begin{lemma}\label{Lemma:Lemma7}
For two $\chi^2$-distributed random variables  with $n$ degrees of freedom, $||X_1||^2$ and $||X_2||^2$ which are projections into orthogonal subspaces of $||X||^2$ with $2n$ degrees of freedom, 
\begin{equation} \label{lemma7}	Pr \left[ 2 \, ||X_1||^2 \ge a' \, ||X||^2 \right] = \epsilon \, , \end{equation}
where $a' = 2(1-\invcdf_{\Beta(a,b)}(\epsilon))$ and $\invcdf_{\Beta(a,b)}(\epsilon)$ is the inverse of the cumulative distribution function of the beta distribution $\Beta(a,b)$ with 
$a=b=\frac{n}{2}$.

\end{lemma}

{\bf Proof:}
It can be shown that (see page 174 of Ref.~\cite{wilks}, and page 71 of Ref.~\cite{Forbes}, noting that the $\chi^2$-distribution is a special case of the Gamma-distribution):
\begin{equation} \frac{||X_1||^2}{||X_1||^2+||X_2||^2}\sim  \Beta\left(\frac{n}{2},\frac{n}{2}\right) \, . \end{equation}
So we can write  
\begin{equation} 
\label{equ:lemma7raw} 
Pr\left[ ||X_1||^2 \ge \invcdf_{\Beta\left(\frac{n}{2},\frac{n}{2}\right)}(1-\epsilon)||X||^2 \right] = \epsilon \, .  
\end{equation}

For the cumulative distribution function ($\cdf$) of the beta distribution, which is the regularized incomplete beta function $I_x(a,b)=\frac{\B(x;a,b)}{\B(a,b)}$ (the incomplete beta function divided by the complete beta function), the following property holds (see Ref.\ \cite{NumRec}, page 178):
\begin{equation} \label{equ:Ix} 
\I_x(a,b) = 1 - \I_{1-x}(b,a) \, . 
\end{equation}
Furthermore, one can show that for $a=b$
\begin{equation} \label{equ:invrelation} \invcdf_{\Beta(a,b)}(1-x) = 
1 - \invcdf_{\Beta(a,b)}(x) \, . 
\end{equation}
Eq.~(\ref{equ:lemma7raw}), together with Eq.~(\ref{equ:invrelation}), and putting $a=b=\frac{n}{2}$, gives Lemma~\ref{Lemma:Lemma7}. $\Box$

Doing the same as above for the upper bound, one also gets
\begin{equation} 
Pr\left[ 2 \, ||X_1||^2 \le b' ||X||^2 \right] \le \epsilon, \end{equation}
where $b'= 2\, \invcdf_{\Beta \left(\frac{n}{2},\frac{n}{2}\right)}(\epsilon)$.

Since Lemma~8 of Ref.~\cite{Leverrier2015} depends on Lemma~7, the former has to be adapted as well:
\begin{lemma}\label{Lemma:Lemma8}
For two vectors $X  \in \mathbb{R}^{4n} $ and $Y \in \mathbb{R}^{4n}$ and $X_1  \in \mathbb{R}^{2n}$, $Y_1  \in \mathbb{R}^{2n}$ projections in a subspace with dimension $n$, the following holds:
\begin{equation} \label{equ:E12} 
Pr\left[ \left| \langle X_1,Y_1 \rangle - \langle X_2,Y_2 \rangle \right| \le \frac{1}{4}(a'-b')(||X||^2+||Y||^2) \right] \ge 1-4\epsilon.\end{equation}
\end{lemma}

{\bf Proof:}
Applying Lemma~\ref{Lemma:Lemma7} to the vectors $X_1+Y_1$ and $X_1-Y_1$, one can write the following inequalities:
\begin{equation}  \label{ineq:1} b' ||X+Y||^2 \le 2 ||X_1+Y_1||^2 \le a' ||X+Y||^2 \end{equation}
\begin{equation} \label{ineq:2} -a' ||X-Y||^2 \le -2 ||X_1-Y_1||^2 \le -b' ||X-Y||^2 
\end{equation}
Note that each of these four inequality holds with probability larger than $1-\epsilon$. We can also write analogous inequalities for $X_2+Y_2$ and $X_2-Y_2$. Also, it can be observed that  
\begin{align}\label{Kolb_sum-difference}
    \langle X_i, Y_i \rangle = \frac{1}{4} ||X_i+Y_i||^2 + \frac{1}{4} ||X_i-Y_i||^2 \, .
\end{align}
Therefore,  adding Eq.~(\ref{ineq:1}) to Eq.~(\ref{ineq:2}) and dividing by 8, one gets a modified version of Lemma~8 of Ref.~\cite{Leverrier2015} 
(with $i \in \{1,2\}$):
\begin{equation} \label{lemma8_1} 
Pr 
\left[ 
\langle X_i,Y_i \rangle 
\leq \frac{1}{4}(b'+a')\langle X,Y \rangle + \frac{1}{8} (a'-b')(||X||^2+||Y||^2) 
\right] 
\geq 1-2\epsilon \, . 
\end{equation}
and
\begin{equation} \label{lemma8_2} 
Pr 
\left[ 
\langle X_i,Y_i \rangle \geq \frac{1}{4}(a'+b')\langle X,Y \rangle
- \frac{1}{8}(a'-b')(||X||^2+||Y||^2)
\right] 
\geq 1-2\epsilon \, . 
\end{equation}
Combining these latter pair of equations one finally gets Lemma~\ref{Lemma:Lemma8}. $\Box$

Now we want to derive the new confidence region for the covariance matrix. Following Ref.~\cite{Leverrier2015}, we consider a thought-experiment where the transmitter (the receiver) splits their generated (measured) {\it real-valued} vectors of length $2n$ in two arbitrary vectors $X_1$ and $X_2$ ($Y_1$ and $Y_2$). (Note that Ref.~\cite{Leverrier2015} used a slightly different notation where the real vector by the transmitter has size $4n$). 
Using the measurement outcomes of the first part of their quantum states, $||X_1||^2$ and $||Y_1||^2$, we can estimate a confidence region for the quantum state describing the remaining part of her quantum state, and vice-versa.

We are in particular interested in the  estimation of the received quadrature variances:
The probability of failure of the estimation is given as in Ref. \cite{Leverrier2015}, where $\mathbb{E}$ denotes the expectation values,
\begin{align} \label{Ebad:Y}
\begin{split} 
p_\text{bad}^{||Y||^2} = Pr \{ & ||Y_1||^2 \ge a \, \, \mathsf{OR} \, \, ||Y_2||^2 \ge a 
 \\
& \, \, \mathsf{OR} \, \, ( ||Y_1||^2 \le a \, \, \mathsf{AND} \, \, \mathbb{E}||Y_2||^2 \ge b ) \\
& \left. \, \mathsf{OR} \, \, ( ||Y_2||^2 \le a \, \, \mathsf{AND} \, \, \mathbb{E}||Y_1||^2 \ge b ) \right\} \leq 6\epsilon \, ,
\end{split} 
\end{align}
where each line's first condition describes the case where the parameter estimation test of the receiver fails, and the second and third lines describe the cases of an incorrect estimation (with parameter estimation test passed).
The same applies, with different bounds, for the
estimation of the covariance term
(see Ref.\ \cite{Leverrier2015}, pages 20~and~24)
\begin{equation} \label{Ebad:XY} 
\begin{split} 
p_{bad}^{\langle X,Y\rangle} =
Pr \left\{ \right.&
\langle X_1,Y_1\rangle \le c 
\, \, \mathsf{OR} \, \, 
\langle X_2,Y_2\rangle \le c \\\
& \, \, \mathsf{OR} \, \, 
( 
\langle X_1,Y_1\rangle \ge c 
\, \, \mathsf{AND} \, \, 
\mathbb{E}\langle X_2,Y_2\rangle \le d
) \\\ 
& \left. \, \, \mathsf{OR} \, \, 
( 
\langle X_2,Y_2\rangle \ge c 
\, \, \mathsf{AND} \, \, 
\mathbb{E}\langle X_1,Y_1\rangle \le d
) \right\}
\leq 6\epsilon  \, .
\end{split} 
\end{equation}
In Eqs.~(\ref{Ebad:Y})-(\ref{Ebad:XY}), there are multiple estimations with $\epsilon$. The probability that one individual estimation fails is given by $6\epsilon$, therefore we put $\epsilon' = \epsilon/6$. In turn, the probability that either the estimation of $\| X \|^2$ or $\langle X , Y \rangle$ fails is given by $6\epsilon$.

The bound for $a$ can be estimated using Lemma~\ref{Lemma:Lemma7}:
\begin{equation} \label{a:equ} a =  \frac{a'(\epsilon')}{2}||X||^2 \, .  \end{equation}
For $b$, one gets from Eq.~(E39) of Ref.~\cite{Leverrier2015}: 
\begin{equation} \label{b:equ} b = a \left( 1+\frac{20}{\epsilon'}e^{-\frac{n}{16}} \right) \, . 
\end{equation}
The factor $c$ can be calculated using Eq.~(\ref{lemma8_2}) 
\begin{equation} \label{c:equ} 
c = \frac{1}{4}\left[ b'(\epsilon')+a'(\epsilon') \right] \langle X,Y\rangle 
+ \frac{1}{8} \left[ b'(\epsilon')-a'(\epsilon') \right] \left( ||X||^2+||Y||^2 \right) \, .
\end{equation}
To compute $d$, we can use Eq.\ (E49) of Ref.\ \cite{Leverrier2015},
\begin{equation} \label{probbad:1} 
\begin{split} 
Pr \left[ ( \mathbb{E}\langle X_2,Y_2\rangle \le d-\delta) \wedge (\langle X_1,Y_1\rangle \ge c) \right] &  \le \frac{d}{\delta} Pr[ \langle X_1,Y_1\rangle -\langle X_2, Y_2\rangle \ge c-d ]\\\ 
& \le \frac{c-\delta}{\delta} Pr\left[ \langle X_1,Y_1\rangle -\langle X_2,Y_2\rangle \ge \delta \right] \\\ 
& \le 
\frac{c}{\delta} Pr\left[ \langle X_1,Y_1\rangle -\langle X_2,Y_2\rangle \ge \delta \right] \\\ & \le \frac{9}{4\epsilon'} Pr\left[ \langle X_1,Y_1\rangle -\langle X_2,Y_2\rangle \ge \delta \right] = \epsilon' \, ,   
\end{split} 
\end{equation} 
whereupon $d=c-\delta$ and $\delta \ge \frac{4c\epsilon'}{9}$ was used. The latter is valid for reasonable parameters, see Ref.~\cite{Leverrier2015} page 25 below Eq.~(E56).
Rewriting Eq.~(\ref{probbad:1}), one gains 
\begin{equation} 
Pr\left[ \langle X_1,Y_1\rangle -\langle X_2,Y_2\rangle \ge \delta \right] = \frac{4}{9}{\epsilon'}^2 \, .  
\end{equation}
With Lemma~\ref{Lemma:Lemma8},
\begin{equation} 
\label{delta} \delta = \frac{1}{4} \left( a''-b'' \right) \left( ||X||^2+||Y||^2 \right) \, ,
\end{equation} 
where $a''$ and $b''$ are defined like $a'$ and $b'$, with their argument $\epsilon'$ replaced by ${\epsilon'}^2/9$.
One then obtains 
\begin{equation} \label{d:equ} 
d = c-\delta = \frac{1}{4}(b'+a')\langle X,Y\rangle +\frac{1}{4}\left( \frac{b'-a'}{2} + b''-a'' \right) \left( ||X||^2+||Y||^2 \right) \, . 
\end{equation}

Finally one gains the bounds by dividing by $n$ :
\begin{equation} \label{equ:sigb} 
y \leq \frac{1}{2n} \, a'(\epsilon') \left(1+\frac{20}{\epsilon'}e^{\frac{-n}{16}} \right)||Y||^2 \, , 
\end{equation}
and 
\begin{equation} \label{equ:sigc} 
z \geq \frac{d}{n} = \frac{1}{2n} \langle X,Y \rangle - \frac{1}{4n} \left( \frac{a'(\epsilon')-b'(\epsilon')}{2} + a'({\epsilon'}^2/9) - b'({\epsilon'}^2/9) \right) \left(||X||^2+||Y||^2\right) \, . 
\end{equation}
\subsection{Assuming Gaussianity}
In the following we derive confidence intervals under the assumption that the random variables are Gaussian. We summarize our findings as
\begin{align}
  \delta_\mathrm{Var}(n,\epsilon) &:= 1 - \frac{1}{2n} \, \mathrm{invcdf}_{\chi^2_{2n}}(\epsilon) \ ,\label{eq:delta_var_gaussian}\\
  \delta_\mathrm{Cov}(n,\epsilon) &:= \frac{1}{2}\left( 1 - \frac{1}{2n} \mathrm{invcdf}_{\chi^2_{2n}}\left( \frac{\epsilon}{2} \right) \right) \ . \label{eq:delta_cov_gaussian}
\end{align}
\subsubsection{Derivation of confidence interval for the variance}
If we assume that the random variables
$q^j_\mathrm{rx}$ and $p^j_\mathrm{rx}$ are Gaussian with variance $y$, then
\begin{align}
\frac{1}{y} \sum_{j=1}^n \left[ \left( q^j_\mathrm{rx} \right)^2 + \left( p^j_\mathrm{rx} \right)^2 \right]
\end{align}
is a chi-square variable with $2n$ degrees of freedom. Therefore we can write 
\begin{align}
Pr \left\{ 
\sum_{j=1}^n \left( q^j_\mathrm{rx} \right)^2 + \left( p^j_\mathrm{rx} \right)^2 < 2n (1 - \delta_\mathrm{Var}) y 
\right\} = \mathrm{cdf}_{\chi^2_{2n}}(2n (1 - \delta_\mathrm{Var})) \, .
\end{align}
That is,
\begin{align}
Pr \left\{ 
\sum_{j=1}^n \left( q^j_\mathrm{rx} \right)^2 + \left( p^j_\mathrm{rx} \right)^2 < 2n (1 - \delta_\mathrm{Var}) y 
\right\} = \epsilon \, ,
\end{align}
for 
\begin{align}
2n (1 - \delta_\mathrm{Var}) = \mathrm{invcdf}_{\chi^2_{2n}}(\epsilon) \, . 
\end{align}

In conclusion, putting
\begin{align}
\delta_\mathrm{Var}(n,\epsilon) := 1 - \frac{1}{2n} \, \mathrm{invcdf}_{\chi^2_{2n}}(\epsilon) \, ,
\end{align}
we obtain 
\begin{align}
Pr \left\{ 
y > \frac{1}{1 - \delta_\mathrm{Var}(n,\epsilon)} \, \frac{1}{n} \sum_{j=1}^n \frac{\left( q^j_\mathrm{rx} \right)^2 + \left( p^j_\mathrm{rx} \right)^2}{2}  
\right\} = \epsilon \, .
\end{align}
For $\delta_\mathrm{Var}(n,\epsilon) \ll 1$ we can use the approximate bound
\begin{align}
Pr \left\{ 
y > (1 + \delta_\mathrm{Var}(n,\epsilon) ) \, \frac{1}{n} \sum_{j=1}^n \frac{\left( q^j_\mathrm{rx} \right)^2 + \left( p^j_\mathrm{rx} \right)^2}{2}  
\right\} = \epsilon \, .
\end{align}
Finally, with our notation this last expression reads
\begin{align}
Pr \left\{ 
y > (1 + \delta_\mathrm{Var}(n,\epsilon) ) \hat y \right\} = \epsilon \, .
\end{align}
\subsubsection{Derivation of confidence interval for the covariance}
For the estimation of the covariance, we need to consider the random variable
\begin{align}\label{sum-difference}
\sum_{j=1}^n q^j_\mathrm{tx} q^j_\mathrm{rx} + p^j_\mathrm{tx} p^j_\mathrm{tx}
& = \frac{1}{4} \sum_{j=1}^n 
\left( q^j_\mathrm{tx} + q^j_\mathrm{rx} \right)^2 
+ \left( p^j_\mathrm{tx} + p^j_\mathrm{rx} \right)^2
- \left( q^j_\mathrm{tx} - q^j_\mathrm{rx} \right)^2
- \left( p^j_\mathrm{tx} - p^j_\mathrm{rx} \right)^2 \, .
\end{align}
Note that the variables 
\begin{align}
& \frac{1}{4} \sum_{j=1}^n \left( q^j_\mathrm{tx} + q^j_\mathrm{rx} \right)^2 + \left( p^j_\mathrm{tx} + p^j_\mathrm{rx} \right)^2 \, , \\
& \frac{1}{4} \sum_{j=1}^n  \left( q^j_\mathrm{tx} - q^j_\mathrm{rx} \right)^2 + \left( p^j_\mathrm{tx} - p^j_\mathrm{rx} \right)^2
\end{align}
are statistically independent and are (not normalized) $\chi^2$ variables with $2n$ degrees of freedom.

Applying the results of previous section we have:
\begin{align}
& Pr \left\{ 
\frac{1}{4} \sum_{j=1}^n \left( q^j_\mathrm{tx} + q^j_\mathrm{rx} \right)^2 + \left( p^j_\mathrm{tx} + p^j_\mathrm{rx} \right)^2 > 2n (1 + \delta_\mathrm{Var}^+) z_+ 
\right\} = 1 - \mathrm{cdf}_{\chi^2_{2n}}(2n (1 + \delta_\mathrm{Var}^+) ) \, , \\
& Pr \left\{ 
\frac{1}{4} \sum_{j=1}^n \left( q^j_\mathrm{tx} - q^j_\mathrm{rx} \right)^2 + \left( p^j_\mathrm{tx} - p^j_\mathrm{rx} \right)^2 < 2n (1 - \delta_\mathrm{Var}^-) z_- 
\right\} = \mathrm{cdf}_{\chi^2_{2n}}(2n (1 - \delta_\mathrm{Var}^-) ) \, ,
\end{align}
where we have put
\begin{align}
z_+ & = \mathbb{E} \left[\frac{1}{4} \left( q^j_\mathrm{tx} + q^j_\mathrm{rx} \right)^2\right] = \mathbb{E}\left[ \frac{1}{4} \left( p^j_\mathrm{tx} + p^j_\mathrm{rx} \right)^2 \right] \, , \\
z_- & = \mathbb{E} \left[\frac{1}{4} \left( q^j_\mathrm{tx} - q^j_\mathrm{rx} \right)^2\right] = \mathbb{E}\left[ \frac{1}{4} \left( p^j_\mathrm{tx} - p^j_\mathrm{rx} \right)^2 \right] \, , \\
z & = z_+ - z_- \, .
\end{align}

For $\delta_\mathrm{Var}^\pm \ll 1$ we can write the approximate expressions:
\begin{align}
& Pr \left\{ 
 z_+ < \frac{1 - \delta_\mathrm{Var}^+}{8n} \sum_{j=1}^n \left( q^j_\mathrm{tx} + q^j_\mathrm{rx} \right)^2 + \left( p^j_\mathrm{tx} + p^j_\mathrm{rx} \right)^2  
\right\} = 1 - \mathrm{cdf}_{\chi^2_{2n}}(2n (1 + \delta_\mathrm{Var}^+) ) \, , \\
& Pr \left\{ 
 z_- > \frac{1 + \delta_\mathrm{Var}^-}{8n} \sum_{j=1}^n \left( q^j_\mathrm{tx} - q^j_\mathrm{rx} \right)^2 + \left( p^j_\mathrm{tx} - p^j_\mathrm{rx} \right)^2 
\right\} = \mathrm{cdf}_{\chi^2_{2n}}(2n (1 - \delta_\mathrm{Var}^-) ) \, ,
\end{align}
which in turn imply
\begin{align}
Pr \left\{ z < 
\frac{1 - \delta_\mathrm{Var}^+}{8n} \sum_{j=1}^n \left( q^j_\mathrm{tx} + q^j_\mathrm{rx} \right)^2 + \left( p^j_\mathrm{tx} + p^j_\mathrm{rx} \right)^2
- \frac{1 + \delta_\mathrm{Var}^-}{8n} \sum_{j=1}^n \left( q^j_\mathrm{tx} - q^j_\mathrm{rx} \right)^2 + \left( p^j_\mathrm{tx} - p^j_\mathrm{rx} \right)^2
\right\} \nonumber \\
\leq 1 - \mathrm{cdf}_{\chi^2_{2n}}(2n (1 + \delta_\mathrm{Var}^+) ) + \mathrm{cdf}_{\chi^2_{2n}}(2n (1 - \delta_\mathrm{Var}^-) ) \, .
\end{align}
The above is equivalent to
\begin{align}
Pr \left\{ z < \hat z -
\frac{\delta_\mathrm{Var}^+}{8n} \sum_{j=1}^n \left( q^j_\mathrm{tx} + q^j_\mathrm{rx} \right)^2 + \left( p^j_\mathrm{tx} + p^j_\mathrm{rx} \right)^2
- \frac{\delta_\mathrm{Var}^-}{8n} \sum_{j=1}^n \left( q^j_\mathrm{tx} - q^j_\mathrm{rx} \right)^2 + \left( p^j_\mathrm{tx} - p^j_\mathrm{rx} \right)^2
\right\} \nonumber \\
\leq 1 - \mathrm{cdf}_{\chi^2_{2n}}(2n (1 + \delta_\mathrm{Var}^+) ) + \mathrm{cdf}_{\chi^2_{2n}}(2n (1 - \delta_\mathrm{Var}^-) ) \, .
\end{align}
Put for simplicity $\delta = \max\{ \delta_\mathrm{Var}^- , \delta_\mathrm{Var}^- \}$:
\begin{align}
Pr \left\{ z < \hat z -
\frac{\delta}{4n} \sum_{j=1}^n \left( q^j_\mathrm{tx} \right)^2 + \left( q^j_\mathrm{rx} \right)^2 + \left( p^j_\mathrm{tx} \right)^2 + \left( p^j_\mathrm{rx} \right)^2
\right\} \nonumber \\
\leq 1 - \mathrm{cdf}_{\chi^2_{2n}}(2n (1 + \delta) ) + \mathrm{cdf}_{\chi^2_{2n}}(2n (1 - \delta) ) \, .
\end{align}
In our notation, 
$\hat x = \frac{1}{2n} \sum_{j=1}^n \left( q^j_\mathrm{tx} \right)^2 + \left( p^j_\mathrm{tx} \right)^2$, 
and 
$\hat y = \frac{1}{2n} \sum_{j=1}^n \left( q^j_\mathrm{rx} \right)^2 + \left( p^j_\mathrm{rx} \right)^2$,
therefore the last expression reads:
\begin{align}
Pr \left\{ z < \hat z -
\frac{\delta}{2} \left( \hat x + \hat y \right) \right\} 
\leq 1 - \mathrm{cdf}_{\chi^2_{2n}}(2n (1 + \delta) ) + \mathrm{cdf}_{\chi^2_{2n}}(2n (1 - \delta) ) \, .
\end{align}
We can further approximate
\begin{align}
Pr \left\{ z < \hat z -
\frac{\delta}{2} \left( \hat x + \hat y \right) \right\} 
\leq 2 \, \mathrm{cdf}_{\chi^2_{2n}}(2n (1 - \delta) ) \, .
\end{align}
Finally, if we put
\begin{align}
\delta = 1 - \frac{1}{2n} \mathrm{invcdf}_{\chi^2_{2n}}\left( \frac{\epsilon}{2} \right) \, ,
\end{align}
the latter reads 
\begin{align}
Pr \left\{ z < \hat z -
\left( 1 - \frac{1}{2n} \mathrm{invcdf}_{\chi^2_{2n}}\left( \frac{\epsilon}{2} \right) \right) \frac{\hat x + \hat y}{2} \right\} 
\leq \epsilon \, .
\end{align}
In conclusion, to have the result stated in our notation, we need to define:
\begin{align}
\delta_\mathrm{Cov}(n,\epsilon) := 
\frac{1}{2}\left( 1 - \frac{1}{2n} \mathrm{invcdf}_{\chi^2_{2n}}\left( \frac{\epsilon}{2} \right) \right) \, ,
\end{align}
which yields the desired relation:
\begin{align}\label{COV00}
Pr \left\{ z < \left( 1 - \delta_\mathrm{Cov}(n,\epsilon) \frac{\hat x + \hat y}{\hat z}\right) \hat z  \right\} 
\leq \epsilon \, .
\end{align}
\subsection{Further optimisation of the confidence interval for covariance estimation}
Let us first show the optimisation under the assumption of Gaussianity. Note that instead of Eq.~(\ref{sum-difference}), we could have used the following identity
\begin{align}
\sum_{j=1}^n q^j_\mathrm{tx} q^j_\mathrm{rx} + p^j_\mathrm{tx} p^j_\mathrm{tx}
& = \frac{1}{4} \sum_{j=1}^n 
\left( \lambda q^j_\mathrm{tx} + \lambda^{-1} q^j_\mathrm{rx} \right)^2 
+ \left( \lambda p^j_\mathrm{tx} + \lambda^{-1} p^j_\mathrm{rx} \right)^2 \nonumber \\
& - \left( \lambda q^j_\mathrm{tx} - \lambda^{-1} q^j_\mathrm{rx} \right)^2
- \left( \lambda p^j_\mathrm{tx} - \lambda^{-1} p^j_\mathrm{rx} \right)^2 \, ,
\end{align}
which holds for any $\lambda >0$.
This would then yield, instead of Eq.~(\ref{COV00}), the following confidence interval for the covariance
\begin{align}\label{COV01}
Pr \left\{ z < \left( 1 - \delta_\mathrm{Cov}(n,\epsilon) \frac{ \lambda^2 \hat x + \lambda^{-2} \hat y}{\hat z}\right) \hat z  \right\} 
\leq \epsilon \, .
\end{align}
The optimal value of $\lambda^2$ is the one that minimizes $\lambda^2 \hat x + \lambda^{-2} \hat y$, i.e., $\lambda^2 = \sqrt{\hat{y}/\hat{x}}$. This finally yields the optimised confidence interval:
\begin{align}\label{COV02}
Pr \left\{ z < \left( 1 - 2 \, \delta_\mathrm{Cov}(n,\epsilon) \frac{ \sqrt{ \hat x \, \hat y } }{\hat z}\right) \hat z  \right\} 
\leq \epsilon \, .
\end{align}

The same argument can be applied also without assuming Gaussianity, and Eq.~(\ref{Kolb_sum-difference}) can be replaced by the identity
\begin{align}
    \langle X_i, Y_i \rangle = \frac{1}{4} || \lambda X_i + \lambda^{-1} Y_i||^2 + \frac{1}{4} || \lambda X_i - \lambda^{-1} Y_i||^2 \, .
\end{align}
In this way, instead of Eq.~(\ref{equ:sigc}) we obtain 
\begin{equation}
z \geq \frac{1}{2n} \langle X,Y \rangle - \frac{1}{4n} \left( \frac{a'(\epsilon')-b'(\epsilon')}{2} + a'({\epsilon'}^2/9) - b'({\epsilon'}^2/9) \right) \left( \lambda ||X||^2 + \lambda^{-1} ||Y||^2\right) \, . 
\end{equation}
Finally, this expression can be optimized by putting $\lambda = ||Y||/||X||$.
This yields
\begin{equation}
z \geq \frac{1}{2n} \langle X,Y \rangle - \frac{1}{2n} \left( \frac{a'(\epsilon')-b'(\epsilon')}{2} + a'({\epsilon'}^2/9) - b'({\epsilon'}^2/9) \right) \, ||X|| \, ||Y|| \, . 
\end{equation}
\section{Secret key calculation}\label{skcalc}
Here we provide details of the various terms in the composable key length calculation~\cite{Leverrier2015, Pirandola2021}. An expression for the lower bound on the composably secure key length is as follows: 
\begin{align}
s_{n}^{\epsilon_{qrng}+\epsilon_h+\epsilon_s+\epsilon_\text{IR}+\epsilon_\mathrm{ent}+\epsilon_\text{PE} + \epsilon_\text{cal}} & \geq
n' \left[ \hat{H}(\bar Y)_{\rho} - I( Y ; E )_{\rho_G} \right] - \text{leak}_\text{IR}(n',\epsilon_{\text{IR}})\nonumber \\
& - \log\left(n'\right)\cdot \sqrt{2n' \log{(2/\epsilon_\mathrm{ent})}}
- \sqrt{n'} \, \Delta_{\mathrm{AEP}}\left(  \frac{p}{3}\epsilon_s^2,d_\text{rx}\right) \nonumber \\
& + \log{\left( p-\frac{p}{3}\epsilon_s^2\right)  }  
+ 2 \log{(\sqrt{2}\epsilon_h)} \, ,
\label{secret-bits-3bis}%
\end{align}
where $n'$ is the total number of exchanged and corrected quantum symbols,  $H(\bar{Y})_{\rho}$ is an entropy estimator, $I(Y;E)_{\rho_G}$ represents the Holevo information (with $\rho_G$ being a Gaussian state having the same covariance matrix as $\rho$), $p$ is the success probability of error correction per frame,   $d_\text{rx}$ is the bit resolution at the receiver, and $\Delta_{\mathrm{AEP}}$ relates the asymptotic limit / infinite number of channel uses with the practical finite-size regime (AEP : asymptotic equipartition property). The various epsilons are summarized in Table~\ref{tab:eps}. Note that $\epsilon_{cal}$, $\epsilon_\text{PE}$, and $\epsilon_\text{qrng}$ do not appear in the RHS of Eq.~(\ref{secret-bits-3bis}) as their role is implicit, for instance, $\epsilon_{cal}$ quantifies the probability that the transmitter or receiver calibration is incorrect. Also $\epsilon_{qrng}$, which represents the overall security parameter of the random numbers at Tx, is limited by the confidence level of the digitization errors in the quantum random number generator (QRNG)~\cite{Gehring2021}. 
\begin{table}[!b]
\centering
\caption{Various security parameters (epsilons) in the key length expression.}
\begin{tabular}{l|l|l}
    \textbf{Name} & \textbf{Description} & Value \\
    \hline
    $\epsilon_{cal}$ & Calibration (both transmitter and receiver) & $10^{-10}$ \\
    $\epsilon_h$ & Hashing function & $10^{-10}$ \\
    $\epsilon_s$ & Smooth min-entropy & $10^{-10}$ \\
    $\epsilon_{ent}$ & Empirical estimate entropy & $10^{-10}$ \\ 
    $\epsilon_\text{PE}$ & Parameter estimation (PE)  & $10^{-10}$ \\
    $\epsilon_\text{IR}$ & Information reconciliation (IR)  & $10^{-12}$ \\
    $\epsilon_\text{qrng}$ & Quantum random number generation  & $2 \times 10^{-6}$ \\
\end{tabular}
\label{tab:eps}
\end{table}
\subsection{Penalty from entropy estimation}
The entropy $H(\bar{Y})_{\rho}$ in Eq 4 from the main paper can be estimated from the empirical frequency 
\begin{align}
f(y_j) = \frac{n'(y_j)}{n'} \, ,
\end{align}
where $n'(y_j)$ is the number of times a specific complex symbol $y_j = q^j_\text{rx} + i p^j_\text{rx}$ is obtained, and $n'$ is the total number of exchanged and corrected quantum symbols. One can define an entropy estimator
\begin{equation}
    \hat H(\bar Y)_{\rho} = - \sum_{j} f(y_j) \log{[f(y_j)]} \, .
\end{equation}
which is linked to $H(\bar{Y})_{\rho}$ by the following inequality \cite{Antos2001, Leverrier2015}:
\begin{align}
H(\bar Y)_{\rho} \geq
\hat H(\bar Y)
- \log\left(n'\right) \, \sqrt{\frac{2 \log{(2/\epsilon_\text{ent})}}{n'}} \, .
\end{align}
This holds true up to a probability smaller than $\epsilon_\mathrm{ent}$.
\subsection{Penalty from the asymptotic equipartition property}
\label{App:AEP}
In Ref.~\cite{Tomamichel2012}, the asymptotic equipartition property bound is proven in Corollary~6.5:
\begin{align}
\frac{1}{n} H^\delta_\mathrm{min}(X^n|E^n) \geq H(X|E) - \frac{\Delta_\mathrm{AEP}(\delta,v)}{\sqrt{n}} \, ,
\end{align}
where
\begin{align}
\Delta_\mathrm{AEP}(\delta,v) := 4 \sqrt{\ell(\delta)} \, \log_2{v} \, ,
\end{align}
and
\begin{align}
v & \leq \sqrt{2^{-H_\mathrm{min}(X|E)}} + \sqrt{2^{H_\mathrm{max}(X|E)}} + 1 \, , \label{eq:AEPv} \\
\ell(\delta) & := - \log_2{\left( 1 - \sqrt{1-\delta^2} \right)} \, .
\end{align}

In the following, we use the fact that $H_\mathrm{min}(X|E)$ is non-negative for our classical-quantum state, a proof of which is given below, in subsection~\ref{skcalc:hminPos}. 
\begin{align}
H_\mathrm{min}(X|E) \geq 0 & \Rightarrow 2^{-H_\mathrm{min}(X|E)} \leq 1 \, , \\
H_\mathrm{max}(X|E) \leq \log_2{2^{2d}} & \Rightarrow \sqrt{2^{H_\mathrm{max}(X|E)}} \leq \sqrt{2^{2d}} = 2^{d} \, .
\end{align}

Using the above relations in Eq.~(\ref{eq:AEPv}) allows us to bound $v$:
\begin{align}
v & \leq \sqrt{2^{-H_\mathrm{min}(X|E)}} + \sqrt{2^{H_\mathrm{max}(X|E)}} + 1 
\leq 2^d + 2 \, .
\end{align}

Now we can easily check that for $d > 1$, 
\begin{align}
\log{(2^d + 2)} < d+1 \, ,
\end{align}
and that
\begin{align}
\ell(\delta) < \log_2{\frac{2}{\delta^2}} \, . 
\end{align}

Putting all together we finally obtain 
\begin{align}
\Delta_\mathrm{AEP}(\delta,v) \leq 4 (d+1) \sqrt{ \log_2{\frac{2}{\delta^2}} } \, ,
\end{align}
which is presented as Eq.~(3) in the main paper. 
\subsubsection{Positivity of min-entropy for a separable state} \label{skcalc:hminPos}
Given two quantum systems $A$ and $B$ in a state $\rho_{AB}$, their smooth-min
entropy is~\cite[Definition 4.1]{Tomamichel2012}%
\begin{equation}
H_{\min}(A|B)_{\rho}:=\max_{\sigma}\sup\{\lambda\in\mathbb{R}:\rho_{AB}%
\leq2^{-\lambda}I_{A}\otimes\sigma_{B}\},
\end{equation}
where $I_{A}$ is the identity operator over $A$ and the maximum is taken over
all sub-normalized states $\sigma_{B}\in\mathcal{S}_{\leq}(\mathcal{H}_{B})$.
Note that%
\begin{equation}
H_{\min}(A|B)_{\rho}\geq\tilde{H}_{\infty}^{\downarrow}(A|B)_{\rho}%
:=\sup\{\lambda\in\mathbb{R}:\rho_{AB}\leq2^{-\lambda}I_{A}\otimes\rho
_{B}\},\label{eq2}%
\end{equation}
where $\tilde{H}_{\infty}^{\downarrow}(A|B)_{\rho}$ can also be defined as the limiting case for $\alpha=\infty$ of a quantum conditional R\'{e}nyi entropy $\tilde{H}_{\alpha}^{\downarrow}(A|B)_{\rho}$ defined in Ref.~\cite[Eq.~(5.18)]{BookTomamichel2015}. 

For separable $\rho_{AB}$, we may write (following Ref.~\cite[Lemma~5.2]%
{BookTomamichel2015})%
\begin{equation}
\rho_{AB}=\sum_{k}p_{k}\theta_{A}^{k}\otimes\rho_{B}^{k}\leq\sum_{k}p_{k}%
I_{A}\otimes\rho_{B}^{k}=I_{A}\otimes\rho_{B}.
\end{equation}
This leads to $\tilde{H}_{\infty}^{\downarrow}(A|B)_{\rho}\geq0$, since we are
left to find the \textit{maximum} value $\lambda_{\max}$ of $\lambda
\in\mathbb{R}$ such that simultaneously satisfies
\begin{equation}
\rho_{AB}\leq I_{A}\otimes\rho_{B},~\rho_{AB}\leq2^{-\lambda}I_{A}\otimes
\rho_{B}.
\end{equation}
Clearly it must be $\lambda_{\max}\geq0$.
\section{Functional blocks}
Below we provide details of the various blocks depicted in Fig. 2 of the main paper. 
\subsection{Quantum randomness generation}
We employed a quantum random number generator (QRNG) based on homodyne measurements of the vacuum state~\cite{Gabriel2010}. As a first step, the QRNG generated uniformly distributed random bit strings by real-time Toeplitz hashing on a field programmable gate array (FPGA). The available min-entropy was bounded by a metrological characterization of the device, compatible with the required security parameter~\cite{Gehring2021}. The uniformly distributed bits were then transformed using the inversion sampling method based on the cumulative distribution function~\cite{Symul2011} to Gaussian-distributed integers with a resolution of 6 bits covering 7 standard deviations. We considered these integers in pairs to form the `quantum data symbol' train, i.e., the complex amplitudes $X = X_I + i X_Q$ of the coherent states used for IQ modulation, as explained in the next subsection. Figures~\ref{fig:s1}(a) and (b) show the histograms for $X_I$ and $X_Q$, respectively, using a total of $10^9$ quantum data symbols. 
\subsection{Modulation (quantum data \&  pilot tone)}
\label{FB:MOD}
\begin{figure*}
\centering
\includegraphics[width=0.98\linewidth]{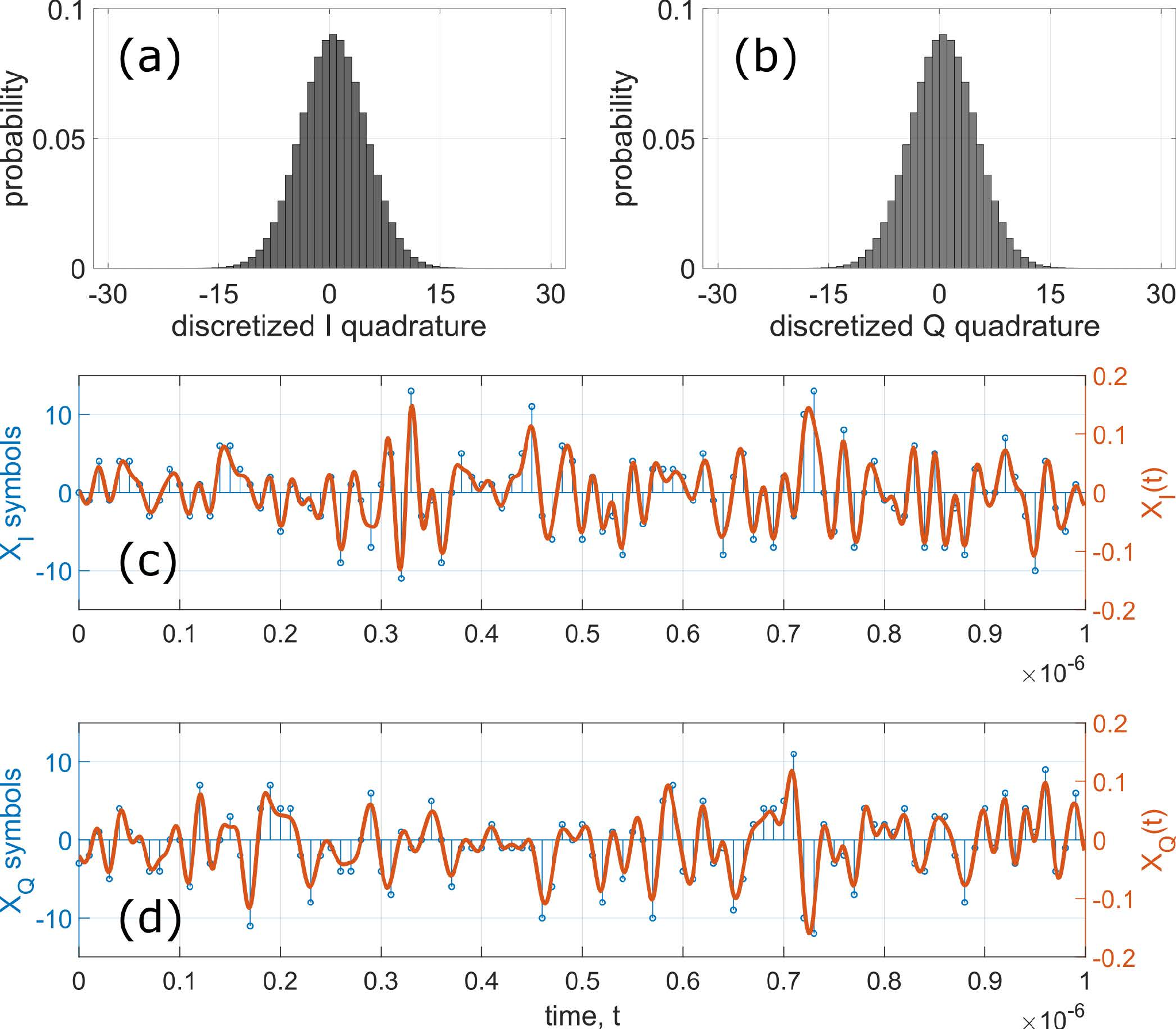}
\caption{\textbf{From raw I and Q symbols to digital waveforms}. (a,b): Histograms of the uniform-to-Gaussian converted symbols $X_I$, $X_Q$. (c,d) The process of creating the digital waveforms involved upsampling (from the symbol bandwidth of 100 MHz to sampling frequency of 1 GHz) while interpolating with a root-raised-cosine filter having a span of 20 symbols. 
\label{fig:s1}}
\end{figure*}

Assuming each quantum data symbol occupying 10 ns in time (quantum data bandwidth $B=100\,$MHz), we created two baseband waveforms $X_I(t)$ and $X_Q(t)$ by upsampling the quantum symbol train to 1\,GSps with interpolation performed using a root-raised-cosine (RRC) filter, characterized by a rolloff factor of 0.2. Note that the variable $t$ is the discretized/sampling time instant with a tick of 1 ns. Figures~\ref{fig:s1}(c) and (d) show the first 100 symbols and the corresponding 1$\mu$s-wide snapshot of the RRC interpolated waveforms. 

For the purpose of sideband modulation and providing a phase reference to the receiver, respectively, we upconverted baseband waveforms to a frequency of $\Omega_u/2\pi = 200\,$MHz and multiplexed a pilot tone in frequency at $\Omega_p/2\pi = 25\,$MHz, as depicted in the inset of Fig. 2 of the main paper. The resulting pair of RF waveforms in the time domain are given by
\begin{align*}
RF1(t) &= X_I(t)\cos(\Omega_u t) - X_Q(t)\sin(\Omega_u t) + A_p\cos(\Omega_p t), \nonumber \\
RF2(t) &= X_I(t)\sin(\Omega_u t) + X_Q(t)\cos(\Omega_u t) + A_p\sin(\Omega_p t).
\label{eq:RF1n2}
\end{align*}
with $A_p$ being a variable for controlling the power in the pilot tone w.r.t. the quantum data band. 
\subsection{Automatic bias control}
In our experiment, we implement the optical single sideband (OSSB) encoding scheme using an IQ modulator, which consists of two nested Mach-Zehnder modulators (MZMs) and a phase modulator (PM). Each of these sub modulators is characterized by a voltage $V_{\pi}$. For the PM, the incoming light wave experiences a $\pi$ phase shift at the output, if the applied DC bias voltage is $V^{\rm PM}_{DC} = V_{\pi}$. The optical transmission through the MZM becomes maximum [minimum] on applying $V^{\rm MZM}_{DC} = k V_{\pi}$ if $k$ is an even [odd] integer. 

Due to environmental drifts etc., the parameter $V_{\pi}$ however exhibits fluctuations, and may require active control. An automatic bias controller continually updates the DC bias voltages by tracking such variations (using a photodiode and a logic circuit for feedback). With appropriate DC biases along with the RF waveforms described in the previous subsection, the optical output shows carrier and sideband suppression, which enables OSSB encoding.  
\subsection{Heterodyne detection, acquisition and demodulation} \label{FB:HAD}
The technique of RF heterodyne detection enables simultaneous measurement of both I and Q quadrature components of the received signal. Heterodyne detection requires that the signal and the local oscillator (LO) have distinct center frequencies. In our experiment, we had a frequency detuning of around $320\,$ MHz, i.e., the signal and LO gave rise to an interference or `beat' signal at the output of the detector; see the right inset of Fig. 2 in the main paper. 

We inserted a low pass filter (cutoff $\approx 365$ MHz) at the detector output for reducing high frequency noise. The filtered output was captured at a sampling rate of 1.0 GSps by an analog-to-digital converter (ADC) inside a fast acquisition card, externally triggered by the AWG. We also used an external 10 MHz clock reference to synchronize the timebase of the AWG with that of the acquisition card. The acquired dataset is divided into `frames', each of length $10^7$ samples. 

The process of reconstructing the transmitted symbols at the receiver required extensive digital signal processing (DSP), which was performed offline on a frame-to-frame basis. In particular, the DSP used a machine learning based approach for optimally tracking the phase between the Tx and Rx lasers (which are free running in our system). The machine learning framework is based on an unscented Kalman filter (UKF), and the primary advantage of doing carrier phase recovery via this approach is the ability to make use of fairly low power pilot tones---in fact, the power of the pilot tone in our experiment is of the same order as the power contained in the quantum signal band~\cite{Chin2020}. Note that at such signal-to-noise ratios, standard phase recovery methods become quite inaccurate. Below we detail some of the DSP routines done on the acquired data frames: 
\begin{enumerate}
    \item \textbf{Pilot frequency estimation:} The pilot frequency was estimated by widely bandpass filtering around the \textit{desired} pilot tone frequency as extracted from a coarsely resolved power spectrum; see the inset on the right in Fig. 2 of the main manuscript. A Hilbert transform was applied to the filtered pilot allowing for extraction of the phase profile by taking its argument. A linear fit yields an estimate of the frequency offset. The procedure was then repeated with a narrower bandpass filter. 
    \item \textbf{Phase estimation:} Using the obtained pilot frequency estimate, we shifted the pilot signal to baseband and downsampled it from the ADC sampling rate of 1.0 GHz to the symbol rate of 100 MHz. We use this signal as the input to the UKF. 
    \item \textbf{Down conversion and phase compensation:} We shifted the quantum data signal also to baseband using the pilot frequency estimate and the known frequency offset between quantum signal and pilot tone ($200-25=175\,$MHz; see Fig. 2 in the main paper) at the Tx. The phase of the pilot tone obtained from the previous stage was used for correcting the phase of the quantum signal. 
    \item \textbf{Timing recovery} By means of cross correlation between the (upsampled version of) reference / Tx symbols from the appropriate frame, and the down-converted quantum signal from the previous stage, we obtained the temporal shift (due to channel propagation and various electronic delays) between the Tx and Rx. The complex quantum signal after synchronization in both quadratures then yields waveforms $Y_I(t)$ and $Y_Q(t)$ corresponding to that frame. 
    \item \textbf{Filtered downsampling:} We used the same RRC filter as explained in subsection~\ref{FB:MOD} to downsample the quantum signal and obtain the final Rx symbols $Y = Y_I + Y_Q$ corresponding to the frame. 
 \end{enumerate}
\section{Evaluation of trusted parameters}
In an ideal world with perfect components, the signal after the quantum channel would not suffer any further loss or incur more noise. As this is not true in practice, the loss and noise owing to the imperfect components that make up the receiver can either be \textit{untrusted}, i.e., fully attributed to Eve, or \textit{trusted} under the assumption that the receiver is physically inaccessible for Eve. We operate in the latter regime, which requires us then to carefully estimate the trusted parameters and incorporate them in the security proof. 

With reference to Fig. 2 from the main paper, the optical loss due to the PC, the (extra, i.e., above 3 dB) loss from the 50/50 BS, and the overall quantum efficiency of the balanced detector contribute to the trusted loss of the receiver. Combining them, we estimated a trusted efficiency $\tau = 0.69$, translating to a loss of 1.6 dB. 

To obtain the trusted noise $t$, we first acquired $10^{10}$ ADC samples with the Rx laser on and the Tx laser off. We calculated the (inverse of the) averaged frequency response from the acquired data samples to create a `whitening' filter. 
\begin{figure*}
\centering
\includegraphics[width=0.9\linewidth]{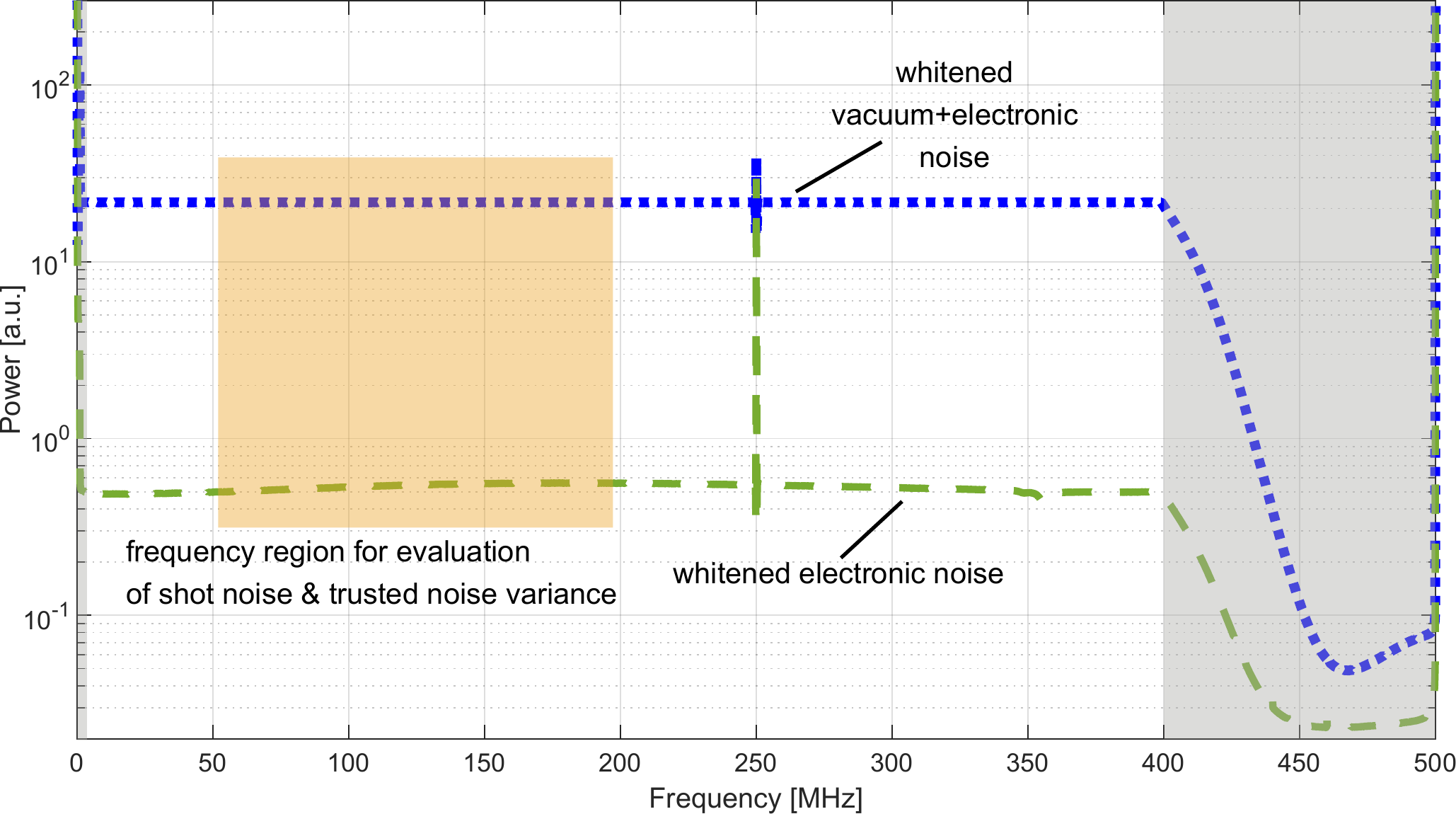}
\caption{\textbf{Receiver calibration for shot noise and trusted noise variances}. 
\label{fig:s2}}
\end{figure*}
Applying such a filter on the original dataset itself yielded a flat response across the entire spectrum, as is illustrated in Fig.~\ref{fig:s2} by the dotted-blue trace. Note we excluded from this whitening process a small low-frequency region (DC to $\lesssim 1\,$MHz) and a high-frequency region ($400-500\,$MHz), depicted by the grey rectangles in Fig.~\ref{fig:s2}, where no components-of-interest are expected. We obtained the dashed-green trace on applying the whitening filter to data samples acquired with both Tx and Rx lasers off. The whitened data samples from both these measurements were subjected to the demodulation procedure outlined in subsection~\ref{FB:HAD}, using the information about the frequency band occupied by the quantum data. 

In this manner, we obtained two sets of $m=10^9$ (complex) symbols over the same band that was occupied by the quantum data signal shown in the inset of Fig. 2. Denoting these two sets by $Y^{(\text{vac})}$ and $Y^{(\text{elec})}$, the shot noise variance is estimated by
\begin{equation}
  \hat{V}_\text{shot} = \text{Var}\left(Y^{(\text{vac})}\right) - \text{Var}\left(Y^{(\text{elec})}\right)\ .
\end{equation}
With $\epsilon_\text{cal}$ as the failure probability for receiver calibration (refer Table~\ref{tab:eps}), one can assign the upper and lower bound within the confidence interval,
\begin{align}
  V_\text{shot}^+ &= (1+\delta_\mathrm{Var}(m,\epsilon_\text{cal}/4))\text{Var}\left(Y^{(\text{vac})}\right) - (1-\delta_\mathrm{Var}(m,\epsilon_\text{cal}/4))\text{Var}\left(Y^{(\text{elec})}\right)\ ,\\
  V_\text{shot}^- &=(1-\delta_\mathrm{Var}(m,\epsilon_\text{cal}/4))\text{Var}\left(Y^{(\text{vac})}\right) - (1+\delta_\mathrm{Var}(m,\epsilon_\text{cal}/4))\text{Var}\left(Y^{(\text{elec})}\right)\ .
\end{align}
An estimate for the mean photon number of the trusted noise $\hat{t}$ is given by
\begin{equation}
\frac{1}{2} \cdot 2\hat{t} + 1 = \frac{\text{Var}\left(Y^{(\text{vac})}\right)}{\text{Var}\left(Y^{(\text{vac})}\right) - \text{Var}\left(Y^{(\text{elec})}\right)} = \frac{1}{1-\frac{\text{Var}\left(Y^{(\text{elec})}\right)}{\text{Var}\left(Y^{(\text{vac})}\right)}}\ ,
\end{equation}
where the factor $\tfrac{1}{2}$ in front of the left hand side of the equation comes from the extra vacuum in heterodyne detection (or equivalently the signal splitting in a phase diverse receiver). Solving for $\hat{t}$ yields
\begin{equation}
\hat{t} = \frac{1}{1-\frac{\text{Var}\left(Y^{(\text{elec})}\right)}{\text{Var}\left(Y^{(\text{vac})}\right)}} - 1\ .
\end{equation}
Since the key fraction is expected to become smaller with less trusted noise (i.e.\ more untrusted noise) we calculate
\begin{equation}
t = \frac{1}{1-\frac{1-\delta_\mathrm{Var}(m,\epsilon_\text{cal}/4)}{1+\delta_\mathrm{Var}(m,\epsilon_\text{cal}/4)}\frac{\text{Var}\left(Y^{(\text{elec})}\right)}{\text{Var}\left(Y^{(\text{vac})}\right)}} - 1\ ,
\end{equation}
as the worst-case trusted noise. 
\section{code efficiency \& reconciliation efficiency}
The information reconciliation (IR) was based on a multi-dimensional (MD) reconciliation scheme and a multi-edge-type low-density-parity-check (MET-LDPC) code of rate 0.19 was used at a SNR of $0.323$. Table~\ref{table:code_dd} shows the degree distribution of this code and its asymptotic performance using density evolution.
 \begin{table*}[ht]
	\caption{The convergence threshold of the MET-LDPC code, denoted by $\sigma^*_\textsubscript{DE}$, is estimated by means of density evolution for a BI-AWGN channel. $\sigma^*_\textsubscript{Sh}$ denotes the threshold at Shannon capacity at which the SNR is specified, while $\beta_\text{Code}^*$ is the asymptotic code efficiency.} 
	\begin{center}\label{table:code_dd}
		\begin{tabular}{|c|c|c|c|c|c|}
			\hline
			$R$ & Degree distribution& $\sigma^*_\textsubscript{DE}$& SNR [dB]&$\sigma^*_\textsubscript{Sh}$&$\beta_\text{Code}^*$\\
                        \hline
            \multirow{2}{*}{0.19}&$\nu(\mathbf{r,x}) =~0.1425\,r_1~x_1^2~x_2^{13} +0.0950\,r_1~x_1^3~x_2^{7}+0.7625\,r_1~x_3~,$&\multirow{2}{*}{$1.77$}&\multirow{2}{*}{$-4.96$}&\multirow{2}{*}{$1.82$}&\multirow{2}{*}{$95.3\%$}\\
            &$\mu(\mathbf{x}) = ~0.0475\,x_1^{12} +~0.5325\,x_2^{3}~x_3^1+~0.230\,x_2^{4}~x_3^1~.$&&&&\\
			\hline
		\end{tabular}
	\end{center}
\end{table*}

The overall MD reconciliation efficiency for a dimension $dim=8$ is given by~\cite{Milicevic2018}: 
\begin{equation*}
    \beta = \beta_\text{Code} \times \beta_\text{Channel} = \beta_\text{Code} \times   \dfrac{C_{dim=8}(s)}{C_{\text{AWGN}}(s)}\, ,
\end{equation*}
where $\beta_\text{Code}$ is the code efficiency in practice (due to the finite length used in the error correction). At the estimated SNR = 0.323, we obtained $\beta_\text{Code} = 94.1$ \% and $\beta_\text{Channel} = 97.4$ \%, yielding an overall reconciliation efficiency of $91.6$ \%.

\bibliography{lib}